\documentclass[12pt]{article}

\newlength{\abstractwidth}
\usepackage{amsmath, nccmath}
\usepackage{amsfonts}
\usepackage{amssymb}
\usepackage{latexsym}
\usepackage{shuffle}

\usepackage{color}
\usepackage{graphicx}
\usepackage{tikz}
\usepackage{dsfont}

\usepackage{tikz}
\usetikzlibrary{calc} 
\usetikzlibrary{patterns,snakes} 
\usetikzlibrary{decorations.pathreplacing} 
\usetikzlibrary{decorations.markings} 
\usetikzlibrary{decorations.pathmorphing} 
\usetikzlibrary{positioning}
\usetikzlibrary{arrows.meta}

\usetikzlibrary{arrows,shapes,positioning}
\usetikzlibrary{decorations.markings}
\usepackage[rightcaption]{sidecap}
\tikzstyle arrowstyle=[scale=1]
\tikzstyle directed=[postaction={decorate,decoration={markings,
    mark=at position .65 with {\arrow[arrowstyle]{stealth}}}}]
\tikzstyle reverse directed=[postaction={decorate,decoration={markings,
    mark=at position .65 with {\arrowreversed[arrowstyle]{stealth};}}}]
\usetikzlibrary{positioning}

\usepackage{hyperref}
\definecolor{darkred}{rgb}{0.8,0.1,0.1}
\hypersetup{colorlinks=true, linkcolor=darkred, citecolor=blue, linktoc=page}

\renewcommand{\thefootnote}{\fnsymbol{footnote}}
\renewcommand{\thanks}[1]{\footnote{#1}}
\newcommand{\starttext}{
\setcounter{footnote}{0}
\setcounter{section}{0}
\renewcommand{\thefootnote}{\arabic{footnote}}}

\newcommand{\bea}{\begin{eqnarray}}
\newcommand{\eea}{\end{eqnarray}}
\newcommand{\be}{\begin{eqnarray}}
\newcommand{\ee}{\end{eqnarray}}
\newcommand{\bma}{\begin{matrix}}
\newcommand{\ema}{\cr\end{matrix}}

\newtheorem{thm}{Theorem}[section]
\newtheorem{lem}[thm]{Lemma}

\def\beq{\begin{equation}}
\def\eeq{\end{equation}}

\def\cC{{\cal C}}

\def\cG{{\cal G}}
\def\cH{{\cal H}}
\def\cI{{\cal I}}
\def\cJ{{\cal J}}

\def\cL{{\cal L}}
\def\cM{{\cal M}}

\def\cO{{\cal O}}

\def\cT{{\cal T}}
\def\cU{{\cal U}}

\def\mA{\mathfrak{A}}
\def\mB{\mathfrak{B}}

\def\mD{\mathfrak{D}}

\def\mJ{\mathfrak{J}}

\def\mw{\mathfrak{w}}

\def\ZZ{{\mathbb Z}}

\def\CC{{\mathbb C}}

\def\Im{{\rm Im \,}}

\def\det{{\rm det \,}}

\def\half{{1\over 2}}
\def\thalf{{\tfrac{1}{2}}}
\def\p{\partial}

\def\a{\alpha}

\def\tet{\vartheta}
\def\ep{\varepsilon}
\def\om{\omega}

\def\oom{\overline{\om}}

\def\no{\nonumber}
\def\sm{\smallskip}

\def\pbx{\p_{\bar x}}
\def\pby{\p_{\bar y}}
\def\pbz{\p _{\bar z}}

\def\bom{\boldsymbol{\om}}
\def\bkappa{\boldsymbol{\kappa}}

\def\bGam{\boldsymbol{\Gamma}}

\begin{document}
\starttext
\setcounter{footnote}{0}

\begin{flushright}
2023 June 14  \\
revised 2025 March 5  \\
UUITP--17/23
\end{flushright}

\bigskip

\begin{center}

{\Large \bf Constructing polylogarithms on}

\vskip 0.1in

{\Large \bf higher-genus Riemann surfaces} 

\vskip 0.4in

{\large Eric D'Hoker$^{(a)}$, Martijn Hidding$^{(b)}$, and Oliver Schlotterer$^{(b)}$}

\vskip 0.1in

 ${}^{(a)}$ {\sl Mani L. Bhaumik Institute for Theoretical Physics}\\
 { \sl Department of Physics and Astronomy }\\
{\sl University of California, Los Angeles, CA 90095, USA}\\

\vskip 0.1in

 ${}^{(b)}$ { \sl Department of Physics and Astronomy,} \\ {\sl Uppsala University, 75108 Uppsala, Sweden}
 
 \vskip 0.1in
 
{\tt dhoker@physics.ucla.edu},  {\tt martijn.hidding@physics.uu.se}, {\tt oliver.schlotterer@physics.uu.se}

\hskip 0.7in

\begin{abstract}
An explicit construction is presented of homotopy-invariant iterated integrals on a Riemann surface of arbitrary genus in terms of a flat connection valued in a freely generated Lie algebra. The integration kernels consist of modular tensors, built from convolutions of the Arakelov Green function and its derivatives with holomorphic Abelian differentials, combined into a flat connection.  Our construction thereby produces explicit formulas for polylogarithms as higher-genus modular tensors. This construction generalizes the elliptic polylogarithms of Brown-Levin, and prompts future investigations into the relation with the function spaces of higher-genus polylogarithms in the work of Enriquez-Zerbini.
\end{abstract}

\end{center}

\newpage

\setcounter{tocdepth}{2} 
\tableofcontents

\newpage

\baselineskip=16pt
\setcounter{equation}{0}
\setcounter{footnote}{0}

\newpage

\section{Introduction}
\label{sec:intro}
\setcounter{equation}{0}

In a variety of research areas in theoretical physics, polylogarithms and related iterated integrals have become almost as widely used as elementary functions. In particular, perturbative computations in quantum field theory and string theory  have  benefitted significantly from the systematic investigation of iterated integrals on the sphere and the torus, namely on Riemann surfaces of genus zero and one. At genus zero, mathematical advances on multiple polylogarithms
have become a driving force behind sophisticated loop calculations in high-energy physics and evaluations of higher-order effective interactions in the low-energy expansion of string theory. At genus one, the construction of elliptic polylogarithms in the mathematics literature  has dramatically increased our computational reach in quantum field theory  and string perturbation theory  and spawned a vibrant collaboration between these two communities. Comprehensive overviews of the literature may be found in reviews and white papers, such as \cite{Bourjaily:2022bwx, Abreu:2022mfk, Blumlein:2022qci, Weinzierl:2022}   for quantum field theory and \cite{Berkovits:2022ivl, Dorigoni:2022iem, DHoker:2022dxx, Mafra:2022wml, DK}  for string theory. 

\sm

A major impetus for the use of polylogarithms and their elliptic analogues is the fact that they span a space of functions which is closed under taking primitives. As a result, integration is rendered completely algorithmic. This property is at the source of the ubiquity of genus-zero polylogarithms \cite{Goncharov:1995, Goncharov:1998kja, Goncharov:2001iea} 
in the study of quantum-field-theory amplitudes \cite{Remiddi:1999ew, Vollinga:2004sn, Goncharov:2010jf, Duhr:2012fh} and string-theory amplitudes  \cite{Broedel:2013tta, Schlotterer:2018zce, Vanhove:2018elu}. Elliptic polylogarithms  were introduced  in  \cite{Beilinson:1994, Levin:2007, BrownLevin}, used to reformulate  Feynman-integral calculations in  \cite{Bloch:2013tra, Adams:2017ejb, Ablinger:2017bjx, Broedel:2017kkb}, and applied to one-loop  string amplitudes in \cite{Broedel:2014vla, DHoker:2015wxz}. 

\sm

The formulation of elliptic polylogarithms crucially hinges on the existence of suitable integration kernels, which were identified in \cite{BrownLevin} and naturally enter genus-one string amplitudes \cite{Dolan:2007eh, Broedel:2014vla, Tsuchiya:2017joo, Gerken:2018}. For the torus, these kernels are usually expressed via Jacobi theta functions, obtained by expanding certain Kronecker-Eisenstein series, and combined into a flat connection. The flatness of the connection guarantees homotopy invariance of the iterated integrals generated by its path-ordered exponential. The translation of Kronecker-Eisenstein kernels from tori to elliptic curves was performed in \cite{Broedel:2017kkb}.

\sm

Higher loop orders of scattering amplitudes in both quantum field theory and string theory involve functions and integrals on higher-genus Riemann surfaces, whose role in string theory dates back to the early days of the subject \cite{Green:1987mn, DHoker:1988pdl,Pol,Witten:2012bh}. In Feynman integrals, both hyperelliptic curves \cite{Huang:2013kh, Georgoudis:2015hca, Doran:2023yzu} and higher-dimensional geometric varieties, such as  Calabi-Yau spaces, have recently been encountered  \cite{Brown:2010bw,  Bourjaily:2018yfy, Duhr:2022pch, Pogel:2022vat, Duhr:2022dxb}.  However, a general and explicit  construction of the functions necessary to describe Feynman integrals and string amplitudes beyond (elliptic) polylogarithms was still missing. For polylogarithms on higher-genus Riemann surfaces, proposals to characterize the function spaces and flat connections have been advanced in the mathematics literature by Enriquez \cite{Enriquez:2011} and Enriquez-Zerbini \cite{Enriquez:2021,Enriquez:2022}, but these have not yet led to tractable expressions for the individual polylogarithms necessary for physics applications. An investigation into the precise relation between these proposals and the construction presented here is relegated to future work.  

\sm
 
In this paper, we shall eliminate this bottleneck for compact Riemann surfaces of arbitrary genus and present a generating series of homotopy-invariant iterated integrals that generalize the polylogarithms and their elliptic analogues to concrete expressions at arbitrary genera. These higher-genus polylogarithms are built out of modular tensors and can be organized to themselves  enjoy tensorial modular transformation properties.

\sm

Our main result here is to provide an explicit proposal for the higher-genus generalization of the integration kernels and flat connection of Brown and Levin. The higher-genus integration kernels in this work share the logarithmic singularities of their genus-one counterparts and the Lie algebra structure of  the formulation of Enriquez and Zerbini \cite{Enriquez:2021}  (see also  \cite{Bernard:1988, Enriquez:2011} for earlier work). While the meromorphic higher-genus connections in the mathematics literature exhibit multi-valuedness \cite{Enriquez:2011}, or poles of arbitrary order \cite{Enriquez:2021} in marked points, our construction features non-meromorphic kernels which reconcile single-valuedness with the presence of at most simple poles.

\sm

Instead of extending the Jacobi theta-function or elliptic-curve description of the Brown-Levin integration kernels used at genus one, our construction on Riemann surfaces of arbitrary genus is driven by the Arakelov Green function \cite{Faltings,Alvarez-Gaume:1986nqf} which, in turn, is built from the prime form \cite{Fay:1973} and Abelian integrals (for a recent account see \cite{DHoker:2017pvk}). We employ convolutions of Arakelov Green functions and their derivatives with holomorphic Abelian differentials to construct higher-genus analogues of the Kronecker-Eisenstein-type integration kernels that were crucial for elliptic polylogarithms. The differential properties of the Arakelov Green function then lead us to identify a flat connection which in turn yields infinite families of homotopy-invariant iterated integrals to be referred to as higher-genus polylogarithms.

\sm

The Arakelov Green function is by now widely used  in string perturbation theory 
\cite{DHoker:2013fcx, DHoker:2014oxd, DHoker:2017pvk, DHoker:2018mys, DHoker:2020tcq}
and has stimulated the construction of tensor-valued functions on the Torelli-space
of compact genus-$h$ Riemann surfaces
\cite{Kawazumi:lecture, Kawazumi:seminar, DHoker:2020uid, Kawazumi:paper}. Our integration kernels for higher-genus polylogarithms enjoy similar tensorial transformation properties under the modular group $Sp(2h,\mathbb Z)$ of compact genus-$h$ Riemann surfaces and are related to the  vector-valued modular forms  investigated by van der Geer and collaborators in \cite{vdG2,vdG3,vdG4}.

\subsection*{Organization}

In section \ref{sec:one}, we review  the construction of polylogarithms at genus zero following Goncharov and at genus one following Brown and Levin, while emphasizing the role played by flat connections. 
Section \ref{sec:higher} provides a summary of the function theory on Riemann surfaces of arbitrary genus $h$ that will be needed here, including the Arakelov Green function. We use these tools to build the $Sp(2h,\ZZ)$  modular tensors needed for the explicit construction of a flat connection  valued in a Lie algebra that is freely generated by $2h$ elements. In section~\ref{sec:polylogs}, we use this connection to construct the promised higher-genus polylogarithms. We discuss their modular properties; provide examples at low order; discuss their meromorphic variants; give evidence for their closure under taking primitives; and present a proposal for higher-genus generalizations of elliptic associators. In section \ref{sec:multi} we generalize the higher-genus connection to the case of multiple marked points on the surface. In section \ref{sec:sepdeg} we consider the behavior of the higher-genus flat connection under separating degenerations and 
recover the Brown-Levin connection at genus one from the degeneration of a genus-two surface.  We conclude and discuss some open directions in section \ref{sec:conc}.


\subsection*{Acknowledgments}
We sincerely thank Benjamin Enriquez and Federico Zerbini for valuable discussions and comments on the first preprint version of the manuscript. ED is grateful to the members of the Theoretical Physics  Group at Uppsala University for their warm hospitality and stimulating environment offered throughout this work. The research of ED is supported in part by NSF grant PHY-22-09700. The research of MH is supported in part by the European Research Council under ERC-STG-804286 UNISCAMP and in part by the Knut and Alice Wallenberg Foundation under grant KAW 2018.0116. The research of OS is supported by the European Research Council under ERC-STG-804286 UNISCAMP.

\newpage

\section{Review of genus-zero and genus-one polylogarithms}
\label{sec:one}
\setcounter{equation}{0}

The construction of homotopy-invariant iterated integrals on a surface of arbitrary genus, including genus zero and genus one, is based on the existence of a flat connection.  We begin by reviewing this well-known construction. A flat connection $\cJ$ on a Riemann surface $\Sigma$, which takes values in a Lie algebra $\cL$, is defined to satisfy the Maurer-Cartan equation, 
\bea
\label{2.MC}
d \cJ - \cJ \wedge \cJ=0
\eea
As a result, the differential equation $d \bGam= \cJ \bGam$ is integrable (and so is $d\bGam' = -  \bGam ' \cJ$). Its solution $\bGam$ takes values in the simply-connected Lie group associated with $\cL$ and is given by the path-ordered exponential along an arbitrary open path $\cC$ between points $z_0,z \in\Sigma$,  
\bea
\label{2.PO1}
\bGam(\cC) = \text{P} \exp \int _\cC \cJ (\cdot) = \text{P} \exp \int _0^1 dt \, J(t)
\eea
We have parametrized the path $\cC$ by $t \in [0,1]$ with $\cC(0)=z_0$ and $\cC(1)=z$ and set $\cJ= J(t) dt$.  The path-ordered exponential is defined by placing $J(t) $ to the left of $J(t')$ for $t>t'$ following physics conventions. Its expansion in powers of $\cJ$ takes the form, 
\bea
\label{2.PO2}
\text{P} \exp \int _0^1 dt \, J(t) =
 1 + \int _0^1 dt_1 \, J(t_1) + \int _0 ^1 dt_1 \int ^{t_1}_0 dt_2 \, J(t_1) J(t_2) + \cdots
 \eea
Flatness of the connection $\cJ$ guarantees that $\bGam(\cC)$ is unchanged under continuous deformations of the path $\cC$, so that $\bGam(\cC)$ depends only on the end-points $z_0$ and $z$. However, $\bGam(\cC)$ may be multiple-valued in $z_0$ and $z$ as these points are taken around non-contractible cycles on $\Sigma$. 
Such iterated integrals will be referred to as \textit{homotopy-invariant}. 

\sm

Polylogarithms on surfaces $\Sigma$ of arbitrary genus are obtained from the path-ordered exponential (\ref{2.PO2}) by extracting the coefficients of independent words in the generators of~$\cL$. Homotopy invariance of $\bGam(\cC)$ 
implies that the resulting polylogarithms are functions of the endpoints $z_0$ and $z$ and  the homotopy class of the path $\cC$, but do not depend on the specific path chosen within a homotopy class. In general, the polylogarithms are multiple valued in $z$ and $z_0$ as these points are taken around a non-trivial homology cycle on $\Sigma$.

\subsection{Genus zero} 

Multiple polylogarithms at genus zero are iterated integrals of rational forms $d z/(z-s)$ with 
$z,s \in  \mathbb C $. They are defined  recursively as follows \cite{Goncharov:2001iea},
\beq
G(s_1,s_2,\cdots,s_n;z ) = \int^z_0 \frac{ d z_1 }{z_1 {-}s_1} \, G(s_2,\cdots,s_n;z_1)
\label{polylog.01}
\eeq
with initial value $G(\emptyset;z ) = 1$ for $s_n \not=0$. When $s_n=0$, an endpoint divergence  is \textit{shuffle-regularized} preserving holomorphicity by setting $G(0;z) = \ln(z)$ (see for example \cite{Panzer:2015ida, Abreu:2022mfk} for pedagogical accounts). The integer  $n\geq 0$  is referred to as the \textit{transcendental weight}. A~generating series for the polylogarithms (\ref{polylog.01}) may be constructed starting from the Knizhnik-Zamolodchikov (KZ) connection $\cJ_{\rm KZ}(z)$ for a Lie algebra $\cL$ that is freely generated by elements $e_1, \cdots, e_m$ associated with the marked points $s_1,\cdots ,s_m$,  
\bea
\label{2.JKZ}
\cJ_{\rm KZ} ( z) = \sum_{i=1}^m { dz \over z-s_i} \, e_i
\eea 
Since $\cJ_{\rm KZ} ( z) $ is a meromorphic $(1,0)$ form in $z \in \hat \CC$, it automatically satisfies the Maurer-Cartan equation (\ref{2.MC}) and is therefore a flat connection, away from the points $s_i$. The path-ordered exponential $\bGam_{\rm KZ} (z)$, produced by the connection $\cJ_{\rm KZ} (z) $ using (\ref{2.PO1}) and (\ref{2.PO2}), is \textit{homotopy-invariant} by construction and depends only on the end-points. Choosing $z_0=0$ by translation invariance and $z_1=z$, we may organize the expansion of the path-ordered exponential in powers of $\cJ_{\rm KZ}$ in terms of the generators $e_1, \cdots, e_m$ (of the universal enveloping algebra of $\cL$), 
\bea
\label{polylog.00}
\text{P} \exp \int^z_0  \cJ_{\rm KZ} ( \cdot)   = 1+ \sum_{\mw}  \mw\,  G(\mw; z)
\eea
The sum runs over all words  $\mw$ with at least one letter, formed out of the alphabet $e_1, \cdots, e_m$, and we identify $G(e_{i_1},e_{i_2},\cdots, e_{i_n} ;z) = G(s_{i_1},s_{i_2},\cdots, s_{i_n};z)$
for $i_1,\cdots,i_n \in \{1,\cdots,m\}$. The construction confirms that every coefficient function $G(\mw;z)$ is a homotopy-invariant iterated integral. Without loss of generality we may set $(s_1,s_2) = (0,1)$ by using the $SL(2,\mathbb C)$ invariance on the sphere, as a result of which these coefficient functions $G(\mw;z)$ reduce to the standard genus-zero polylogarithms in $m{-}1$ variables \cite{Goncharov:1995, Goncharov:1998kja, Goncharov:2001iea}.

\subsection{Genus one} 
\label{sec:blpoly}

At genus zero, the coefficient of each Lie-algebra generator $e_i$ in the connection (\ref{2.JKZ}) is a single-valued meromorphic $(1,0)$-form with simple poles (as opposed to higher-order poles). On a compact Riemann surface of genus $h \geq 1$, however, it is not possible to maintain these properties simultaneously without introducing additional marked points. Instead, the available options are as follows.
\begin{enumerate}
\itemsep -0.05in
\item a single-valued but non-meromorphic connection with at most simple poles;
\item a meromorphic but not single-valued connection with at most simple poles; 
\item a meromorphic and single-valued connection with a single pole of higher order, or poles of lower order distributed over multiple marked points.
\end{enumerate}
The Brown-Levin construction of elliptic polylogarithms via iterated integrals \cite{BrownLevin} follows option 1, and will be briefly reviewed below.\footnote{An alternative construction of elliptic polylogarithms given in the same Brown-Levin reference \cite{BrownLevin} relies on certain averages of genus-zero polylogarithms which preserve meromorphicity. Throughout this work, we use the term \textit{Brown-Levin (elliptic) polylogarithms} to refer to the non-meromorphic
iterated integrals in \cite{BrownLevin} and not to the meromorphic functions obtained from the averaging procedure of the reference.} It will be generalized  to higher genus in the next subsection following option~1. The constructions at 
genus one following options~1 and~2 will be related at the end of this section, while their relation with option 3 will be relegated to future work.

\subsubsection{The Brown-Levin construction} 

A genus-one surface $\Sigma$ with modulus $\tau$ in the Poincar\'e upper half plane may be represented by $\Sigma = \CC/ (\ZZ+ \tau \ZZ)$, and parametrized by local complex coordinates $z, \bar z$ subject to identifications $z \equiv z+1$ and $z \equiv z+\tau$.  The Brown-Levin connection $\cJ_{\rm BL}(z|\tau)$ takes values in the Lie algebra $\cL$ generated freely by elements $a,b$ and is given as follows,
\bea
\label{polylog.06a}
\cJ_{\rm BL}( z|\tau) = {\pi \over \Im \tau} \, \big ( d  z {-} d \bar z\big )   \, b + d z \, \bigg( a
+ \sum_{n=1}^{\infty} f^{(n)}(z| \tau) \, {\rm ad}_b^n(a) \bigg)
\eea
where ${\rm ad}_b(\cdot) = [b, \cdot]$. Flatness of the connection, namely $d \cJ_{\rm BL} - \cJ_{\rm BL} \wedge \cJ_{\rm BL}=0$, requires the following relations between the coefficient functions $f^{(n)}(z|\tau)$,\footnote{As we shall see below, the relation for $n=1$ actually holds up to a $\delta$-function, $\pbz f^{(1)} (z) = \pi \delta (z) - \pi /\Im \tau$, so that the corresponding relation (\ref{2.fw}) holds for $z \not=0$, as does the flatness condition of $\cJ_{\rm BL}$.  Throughout, we shall set $d^2z = { i \over 2} dz \wedge d\bar z$ and normalize the $\delta$-function by $\int_\Sigma d^2 z \, \delta (z)=1$.}
\bea
\label{2.fw}
\pbz f^{(n)} (z|\tau) = - {\pi \over \Im \tau} \, f^{(n-1)}(z|\tau) \hskip 1in f^{(0)}(z|\tau)=1
\eea
The functions $f^{(n)}(z|\tau)$ may be constructed in different but equivalent ways. Following Brown and Levin, they are given by expanding the doubly-periodic Kronecker-Eisenstein series $\Omega(z,\alpha|\tau)$ in powers of an auxiliary parameter $\a \in \CC$, 
\bea
\Omega(z,\alpha| \tau) = \exp \bigg( 2\pi i  \alpha\, \frac{ \Im z }{ \Im \tau} \bigg) \,
\frac{ \tet_1'(0|\tau) \tet_1(z{+}\alpha|\tau) }{\tet_1(z|\tau) \tet_1(\alpha|\tau)}
=  \sum_{n=0}^\infty \alpha^{n-1} f^{(n)}(z|\tau) 
\label{polylog.04}
\eea
The relations (\ref{2.fw}) immediately result from the following identity for $\Omega(z,\alpha|\tau)$ for $z\not=0$, 
\bea
\pbz \Omega(z,\alpha|\tau) = - \frac{ \pi\,\alpha }{\Im \tau} \,  \Omega(z,\alpha|\tau)  
\label{polylog.07}  
\eea
One may obtain  the connection $\cJ_{\rm BL}$ in (\ref{polylog.06a}) by a formal substitution $\a \to {\rm ad}_b$ as follows,
\bea
\label{polylog.06b}
\cJ_{\rm BL}(z|\tau) = \frac{ \pi }{\Im \tau} \, (  d  z {-} d \bar z ) \, b 
+ d z \, {\rm ad}_b \, \Omega \big (z, {\rm ad}_b|\tau \big ) \, a
\eea
The factor ${\rm ad}_b$ to the left of $\Omega$ ensures the cancellation of the pole that $\Omega(z,\alpha|\tau)$ has in $\a$.

\subsubsection{Alternative construction via convolutions of Green functions}

An alternative construction of the functions $f^{(n)}(z|\tau)$, and the one that will generalize to higher genus, is in terms of the scalar Green function $g(z|\tau)$ on $\Sigma$, which is defined by,
\bea
\label{2.green}
\pbz \p_z \, g(z|\tau) = - \pi \delta (z) + { \pi \over \Im \tau} \hskip 1in \int_\Sigma d^2 z \, g(z|\tau) =0
\eea
and may expressed in terms of $\tet$-functions and the Dedekind eta-function $\eta$ as follows,
\bea
g(z|\tau) = - \ln \left | { \tet_1(z|\tau) \over \eta(\tau) } \right |^2 -  \pi { (z {-}\bar z)^2 \over 2 \, \Im \tau}
\eea
Furthermore, we define two-dimensional convolutions of $g$ recursively as follows,
\bea
\label{2.concat}
g_{n+1}(z|\tau) = \int _\Sigma { d^2 x \over \Im \tau}  \, g(z{-}x|\tau) \,  g_n(x|\tau) \hskip 1in g_1(x|\tau) = g(x|\tau)
\eea
In terms of these convolutions $g_n(z|\tau)$ the integration kernels $f^{(n)}(z|\tau)$ are given by, 
\bea
\label{2.fg}
f^{(n)}(z|\tau) = - \p_z ^n g_n(z|\tau)
\eea
and may thus also be defined recursively by convolutions over $\Sigma$ \cite{Gerken:2018}, 
\bea
 f^{(n)}(z|\tau ) = - \int_{\Sigma} \frac{ d^2 x}{\Im \tau} \, f^{(1)}(z{-}x|\tau ) \,f^{(n-1)}(x|\tau) 
  \hskip 1in n\geq 2
\label{polylog.17}
\eea
We note that, in co-moving coordinates $u,v\in [0,1]$ with $z=u\tau {+} v$, the non-holomorphic prefactor in the definition (\ref{polylog.04}) of $\Omega (z,\a|\tau)$ becomes $e^{2 \pi i \a u}$ so that, for fixed $u,v$,  the functions $f^{(n)}(u\tau{+}v|\tau)$ are holomorphic in the modulus $\tau$.

\subsubsection{Modular properties of the Brown-Levin construction}
\label{sec:ctaud}

Under a modular transformation on the modulus $\tau,z$, and $\a$ given by,
\bea
\tau \to \tilde \tau = { A \tau +B \over C \tau +D}
\hskip 0.8in
z \to \tilde z = { z \over C \tau +D} 
\hskip 0.8in
\a \to \tilde \a = { \a \over C \tau +D}
\label{h1mod}
\eea
where $A,B,C,D\in \ZZ$ with $AD-BC=1$, the Kronecker-Eisenstein series $\Omega$ and the functions $f^{(n)}$ in (\ref{polylog.04}) transform as modular forms of weight $(1,0)$ and $(n,0)$, respectively,
\bea
\Omega ( \tilde z, \tilde \a| \tilde \tau) & = & (C \tau + D) \Omega(z,\a|\tau)
\no \\
f^{(n)}(\tilde z|\tilde \tau) & = & (C \tau+D)^n f^{(n)}(z|\tau)
\eea
These transformation properties may be readily established by using (\ref{polylog.04}) and the transformation properties of the Jacobi $\tet$-function, 
\bea
\tet _1(\tilde z, \tilde \a|\tilde \tau) 
= \ep \, (C \tau +D)^\half e^{i\pi  C z^2/(C\tau+D)} \tet_1(z|\tau) \hskip 0.8in \ep^8=1
\label{tettrfm}
\eea
or the modular invariance of the functions $g_n(z|\tau)$ along with the relation (\ref{2.fg}). 
The modular properties of the Brown-Levin connection and polylogarithms 
are most transparent by assigning the following transformation law to the 
generators $a,b$  in (\ref{polylog.06b}),
\bea
a \to \tilde a = (C \tau +D) a + 2\pi i C b \hskip 1in b \to \tilde b = { b \over C\tau+D}
\label{goodmod}
\eea
This choice renders the flat connection $\cJ_\text{BL}$
modular invariant under (\ref{h1mod}). The extra contribution $2\pi i C b$ to 
$\tilde a$ in (\ref{goodmod}) is engineered to compensate 
the transformation of the first term in the expression (\ref{polylog.06b})
for the connection
\bea
{\pi \, d\tilde z \over \Im  \tilde \tau}  \, \tilde b  = { C \bar \tau +D \over C \tau +D} \, {\pi \,  dz \over \Im \tau}   \, b 
\eea

\subsubsection{Homotopy-invariant iterated integrals}

Homotopy-invariant iterated integrals on a genus-one surface are constructed by expanding  the path-ordered exponential in terms of words in the (rather frugal) alphabet $a,b$ as follows,
\bea
 \label{polylog.08} 
\text{P} \exp \int^z_0 \cJ_{\rm BL}( \cdot |\tau)  = 1+ 
 \sum_{\mw } \mw \, \Gamma(\mw; z|\tau)
\eea
The Brown-Levin connection $\cJ_{\rm BL}$ can be found in (\ref{polylog.06b}), and the sum runs over all words  $\mw$ with at least one letter, formed out of the alphabet $a,b$.  The construction guarantees that every coefficient function $\Gamma (\mw;z|\tau)$ is a homotopy-invariant iterated integral. These functions were dubbed elliptic polylogarithms in \cite{BrownLevin}. 

\sm

In this construction, the requirement of  doubly-periodicity  introduces non-holomorphicity into the functions $f^{(n)}(z|\tau)$, into the connection $\cJ_{\rm BL}$, and into the elliptic polylogarithms in (\ref{polylog.08}). Still, any $\bar z$-dependence of $\Gamma(\mw; z|\tau)$ occurs only via polynomials  in $2\pi i ( \Im z)/ \Im \tau$, so that the key structure is carried essentially by meromorphic iterated integrals.

\subsubsection{Meromorphic variant}

Given the meromorphic counterparts of the doubly-periodic Kronecker-Eisenstein series $\Omega(z,\alpha|\tau)$ and its expansion coefficients $f^{(n)}(z|\tau)$ of (\ref{polylog.04}),\footnote{The meromorphic functions $g^{(n)}(z|\tau)$ in the expansion given in (\ref{polylog.10}) are not to be confused with the real-analytic convolutions $g_n(z|\tau)$ of the scalar Green function on the torus defined in (\ref{2.concat}). Both notations have, by now, become standard for historical reasons and can be distinguished through the placement of $n \in \mathbb N_0$ in the superscript and in parenthesis in case of (\ref{polylog.10}).}
\beq
\frac{ \tet_1'(0|\tau ) \tet_1(z{+}\alpha |\tau) }{\tet_1(z|\tau ) \tet_1(\alpha|\tau )} =
  \sum_{n=0}^\infty \alpha^{n-1}  g ^{(n)}(z|\tau) 
\label{polylog.10} 
\eeq
a meromorphic variant  of the Brown-Levin polylogarithms (\ref{polylog.08}) may be recursively defined as follows \cite{Broedel:2017kkb},
\beq
\tilde \Gamma\big( \smallmatrix n_1 &n_2 &\cdots &n_r \\ a_1 &a_2 &\cdots &a_r \endsmallmatrix;z|\tau \big)
= \int^z_0 \! d z_1 \, g ^{(n_1)}(z_1 {-}a_1|\tau) \,
\tilde \Gamma\big( \smallmatrix n_2 &\cdots &n_r \\ a_2 &\cdots &a_r \endsmallmatrix;z_1 |\tau \big)
\label{polylog.11} 
\eeq
with $\tilde \Gamma(\emptyset;z|\tau)=1$ and $r,n_i \in \mathbb N_0$.  These meromorphic elliptic polylogarithms coincide with the formulation via $\Gamma\big( \smallmatrix n_1 &n_2 &\cdots &n_r \\ a_1 &a_2 &\cdots &a_r \endsmallmatrix;z|\tau \big)$ in \cite{Broedel:2014vla} on the real line and exhibit a closer analogy to the recursive definition (\ref{polylog.01}) at genus zero than the $\Gamma(\mw; z|\tau)$ in (\ref{polylog.08}). However, the meromorphic integration kernels $g^{(n)}(z|\tau)$ in (\ref{polylog.10}) such as
\beq
g ^{(0)}(z|\tau)  = 1
\hskip 1in
g ^{(1)}(z|\tau)  = \partial_z \ln \tet_1(z|\tau) =  f ^{(1)}(z|\tau)  - 2\pi i\, \frac{ \Im z}{\Im \tau}
\label{exft01}
\eeq
which enter the construction (\ref{polylog.11}) of $\tilde \Gamma$ are generically multiple-valued on the torus, and thus more properly live on the universal covering space, which is $\CC$.  
 
 \sm
 
Note that the Brown-Levin polylogarithms (\ref{polylog.08}) associated with
words $\mw \rightarrow a b \cdots b$ reduce to a single integral over the
meromorphic kernels (\ref{polylog.10}), for instance
\begin{align}
\Gamma(ab;z|\tau) &= 
 \int^z_0 d t \, \bigg(  2\pi i  \, \frac{ \Im t}{\Im \tau} - f^{(1)}(t|\tau)   \bigg)
 = 
- \int^z_0 d t \,  g^{(1)}(t|\tau)  
 = - \tilde \Gamma\big( \smallmatrix 1 \\ 0 \endsmallmatrix;z|\tau \big)
 \label{polylog.12a} 
\end{align}
(see \cite{Broedel:2014vla, Broedel:2018iwv}
for the regularization of endpoint divergences)
and more generally 
\begin{align}
\Gamma(a\underbrace{b\cdots b}_{n};z|\tau) &= (-1)^n 
 \int^z_0 d t \, \sum_{j=0}^n \frac{1}{j!} \, \bigg( {-} 2\pi i  \, \frac{ \Im t}{\Im \tau} \bigg)^j f^{(n-j)}(t|\tau)  
  \notag \\
  &=(-1)^n 
 \int^z_0 d t \,  g^{(n)}(t|\tau)  
 = (-1)^n \tilde \Gamma\big( \smallmatrix n \\ 0 \endsmallmatrix;z|\tau \big)
 \label{polylog.12b} 
\end{align}
These examples illustrate that the Brown-Levin polylogarithms in (\ref{polylog.08}) are
$\mathbb Q[2\pi i \frac{ \Im z }{\Im \tau}]$-linear combinations
of the meromorphic ones in (\ref{polylog.11}).

\newpage

\section{A flat connection at higher genus}
\label{sec:higher}
\setcounter{equation}{0}

This section is dedicated to the construction of a flat connection  which generalizes the Brown-Levin flat connection $\cJ_{\rm BL}$ in (\ref{polylog.06b}) to higher genus. We begin by introducing some functions and forms on Riemann surfaces of arbitrary genus that will provide the key ingredients in our construction. Further background material on Riemann surfaces and their function theory, including $\tet$-functions, may be found in \cite{Fay:1973, Mumford:1983, DHoker:1988pdl}.

\subsection{Basics}

The topology of a  compact Riemann surface $\Sigma$ without boundary is specified by its genus~$h$. The homology group $H_1(\Sigma , \ZZ)$ is isomorphic to $\ZZ^{2h}$ and supports an anti-symmetric non-degenerate intersection pairing  denoted by $\mJ$.   A canonical homology basis of cycles $\mA_I$ and $\mB_J$ with $I,J=1,\cdots, h$ has  symplectic intersection matrix $\mJ(\mA_I, \mB_J)  = - \mJ(\mB_J, \mA_I)  =  \delta _{IJ}$, and $\mJ(\mA_I, \mA_J)  = \mJ(\mB_I, \mB_J)  =  0$, as illustrated in Figure 1 for genus two.

\begin{figure}[htb]
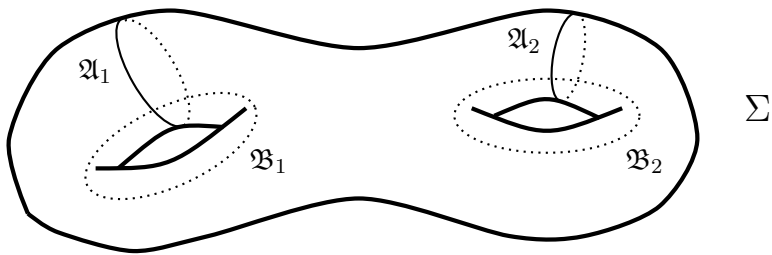

\begin{center}
\tikzpicture[scale=1]
\scope[xshift=-5cm,yshift=0cm]
\draw[ultra thick] plot [smooth] coordinates {(0.6,0.6) (0.35,1.6) (1,2.8) (2.5, 3.3)  (5,2.8) (7.5, 3.3) (9,2.8) (9.5,1.6) (9,0.7) (8,0.3) (7, 0.3) (5,0.8) (3, 0.3) (2,0.1) (1,0.35) (0.6,0.6) };
\draw[ultra thick] (1.5,1.2) .. controls (2.5, 1.2) .. (3.5,2);
\draw[ultra thick] (1.8,1.2) .. controls (2.5, 1.8) .. (3.2,1.75);
\draw[ultra thick] (6.5,2) .. controls (7.5, 1.6) .. (8.5,2);
\draw[ultra thick] (6.8,1.9) .. controls (7.5, 2.2) .. (8.2,1.87);
\draw[thick, dotted,  rotate=25] (2.9,0.3)  ellipse (35pt and 15pt);
\draw[thick, dotted,  rotate=00] (7.5,1.9)  ellipse (35pt and 14pt);
\draw[thick, dotted,  rotate=30] (3.2,0.2)  arc (-90:90:0.3 and 0.8);
\draw[thick,  rotate=30] (3.2,1.84)  arc (90:270:0.3 and 0.82);
\draw[thick, dotted,  rotate=-10] (7.2,3.4)  arc (-90:90:0.2 and 0.6);
\draw[thick,  rotate=-10] (7.2,4.58)  arc (90:270:0.2 and 0.59);
\draw (1.5,2.5) node{\small $\mA_1$};
\draw (7.2,2.9) node{\small $\mA_2$};
\draw (3.8,1.3) node{\small $\mB_1$};
\draw (8.8,1.3) node{\small $\mB_2$};
\draw (10.3,2) node{{\large $\Sigma$}};
\endscope
\endtikzpicture
\caption{A choice of canonical homology basis on a compact genus-two Riemann surface $\Sigma$.}
\end{center}
\label{fig:1}
\end{figure}

\vskip -0.2in

\noindent
A canonical basis of holomorphic Abelian differentials $\bom_I$ may be normalized  on $\mA$-cycles,\footnote{Throughout, differential forms are denoted in boldface, while their component functions in local complex coordinates $z, \bar z$ on $\Sigma$ are denoted by the same letter in normal font, such as for example 
$\bom_I= \omega_I(z) d z$. The coordinate volume form on $\Sigma$ is $d^2 z = \frac{i}{2} d z \wedge d \bar z$ and the $\delta$-function is normalized by $\int _\Sigma d^2 z \, \delta (z,w) =1$
for any $w \in \Sigma$. Finally, repeated pairs of identical indices are to be summed following the Einstein convention so that, for example, we set $Y^{IJ} \bom_J = \sum _{J=1}^h Y^{IJ} \bom_J$.}
\bea
\label{omnorm}
\oint _{\mA_I} \bom_J = \delta _{IJ} 
\hskip 1in 
\oint _{\mB_I} \bom_J = \Omega _{IJ} 
\eea
The complex variables $\Omega _{IJ}$ denote the components of the period matrix $\Omega$ of the surface $\Sigma$.  By the Riemann relations, $\Omega$ is symmetric, and has positive definite imaginary part,
\bea
\Omega ^t = \Omega 
\hskip 1in 
Y = \Im \Omega >0
\eea
The matrix $Y_{IJ}= \Im \Omega _{IJ}$ and its inverse $Y^{IJ} = \big ( (\Im \Omega)^{-1} \big )^{IJ}$ may be used to raise and lower indices as follows,
\bea
\bom^I = Y^{IJ} \bom_J \hskip 0.8in \bar \bom ^I = Y^{IJ} \bar \bom_J \hskip 0.8in Y^{IK} Y_{KJ} = \delta ^I_J
\label{raiseind}
\eea 
In particular, it will be useful to express the Riemann bilinear relations as follows,
\bea
{i \over 2} \int _\Sigma \bom_I \wedge \bar \bom ^J = \delta _I^J
\eea
The choice of canonical $\mA$ and $\mB$-cycles is not unique: a new  canonical basis $\tilde \mA$ and $\tilde  \mB$ is obtained by applying a modular transformation $M \in Sp(2h,\ZZ)$,  such that $M$ satisfies $M^t \mJ M= \mJ$, and the column matrices of cycles are transformed as follows,
\bea
\label{3.mod1}
\left ( \begin{matrix} \tilde \mB \cr \tilde \mA \cr \end{matrix} \right )= 
\left ( \begin{matrix} A & B \cr C & D \cr \end{matrix} \right )  
\left ( \begin{matrix} \mB \cr \mA \cr \end{matrix} \right )
\hskip 0.8in M = \left ( \begin{matrix} A & B \cr C & D \cr \end{matrix} \right ) 
\hskip 0.8in \mJ = \left ( \begin{matrix} 0 & -I \cr I & 0 \cr \end{matrix} \right ) 
\eea
Under a modular transformation $M$, the row matrix $\bom$ of holomorphic Abelian differentials $\bom_I$,  the  period matrix $\Omega$, and its imaginary part $Y$, transform as follows, 
\bea
\label{3.mod2}
\tilde \bom  & = & \bom \, (C \Omega +D) ^{-1} 
\no \\
\tilde \Omega   & = & (A \Omega +B) (C\Omega +D)^{-1}
\no \\  
\tilde Y & = & \left (\bar \Omega C^t + D^t \right )^{-1} \, Y \, (C\Omega +D)^{-1}
\eea
where we have denoted transformed quantities with a tilde as in the discussion of the genus-one case in  section \ref{sec:ctaud}.  

\sm

The moduli space of compact Riemann surfaces of  genus $h$  will be denoted by $\cM_h$. The moduli space $\cM_h$ for $h=1,2,3$ may be identified with $\cH_h/Sp(2h,\ZZ)$ provided we remove from the Siegel upper half space $\cH_h$  for $h=2,3$ all elements which correspond to disconnected surfaces, and take into account the effect of automorphisms including the involution on the hyper-elliptic locus for $h=3$. For $h\geq 4$, the moduli space $\cM_h$ is a complex co-dimension $\half (h-2)(h-3)$ subspace of $\cH_h/Sp(2h,\ZZ)$ known as the Schottky locus.

\subsection{The Arakelov Green function} 

The Arakelov Green function $\cG(x,y|\Omega)$ on $\Sigma \times \Sigma$ generalizes the Green function $g(z|\tau)$ which was defined at genus one in (\ref{2.green}), and is defined by,
\bea
\label{3.Arak}
\pbx \p_x \cG(x,y|\Omega) = - \pi \delta (x,y) + \pi \kappa (x) \hskip 1in \int _\Sigma \bkappa(x) \cG(x,y|\Omega)=0
\eea
where $\kappa$ is given by the pull-back to $\Sigma$ under the Abel map of the unique translation invariant K\"ahler form on the Jacobian variety $J(\Sigma)= \CC^h / (\ZZ^h + \Omega \ZZ^h)$, normalized to unit volume,\footnote{A recent account of the Arakelov Green function and its properties needed here may for instance be found in \cite{DHoker:2017pvk}.}
\bea
\bkappa = { i \over 2h} \bom_I \wedge \bar \bom^I = \kappa (z) \, d^2 z \hskip 1in \int _\Sigma \bkappa=1
\label{polylog.24}
\eea 
Here and throughout the rest of this work, we shall suppress the dependence 
on the period matrix $\Omega$ unless otherwise indicated.
Both $\bkappa$ and $\cG(x,y)$ are conformally invariant. An explicit formula for $\cG(x,y)$ may be given in terms of the non-conformally invariant string Green function $G(x,y)$ as follows, 
\bea
\cG(x,y) = G(x,y) - \gamma (x) -\gamma (y) + \gamma_0 
\label{polylog.25}
\eea
where $\gamma(x)$ and $\gamma _0$ are given by,
\bea
\gamma (x) = \int _\Sigma \bkappa(z) G(x,z) 
\hskip 1in
\gamma_0 = \int _\Sigma \bkappa \gamma 
\label{polylog.26}
\eea
The string Green function is given in terms of the prime form $E(x,y)$ by,\footnote{Here, it is understood that the prime form, and thus the string Green function, are defined in a suitable  fundamental domain for $\Sigma$ in the universal covering space of $\Sigma$, and that the integrals in (\ref{polylog.26}) are to be carried out in that same domain \cite{DHoker:2017pvk}. }
\bea
G(x,y) = {-} \log | E(x,y) |^2 + 2\pi  \left ( \Im \int^x_y \!  \bom_I \right ) \left (   \Im \int^x_y \!  \bom^I \right )
\label{polylog.23}
\eea
see Appendix \ref{sec:pform} for the definition of the prime form
in terms of theta functions. The following double derivatives, 
\bea
\label{3.Arak2}
\p_x \p_y \cG(x,y) & = & - \p_x \p_y \ln E(x,y) + \pi \, \om_I(x) \, \om^I(y)
\no \\
\p_x \pby \cG(x,y) & = & \pi \, \delta(x,y) - \pi \, \om_I(x) \, \bar \om^I(y)
\eea
will prove useful in the sequel.

\subsection{Convolution of Arakelov Green functions and modular tensors} 
\label{sec:gtensor}

Convolutions involving the Arakelov Green function and various integration measures were used to construct a variety of modular tensors in \cite{Kawazumi:lecture, Kawazumi:seminar,DHoker:2020uid, Kawazumi:paper}. Here we shall extend this library of modular tensors  by including convolutions that involve not only Arakelov Green functions (as was the case in \cite{DHoker:2020uid}) but also derivatives on Arakelov Green functions and holomorphic Abelian differentials. The building blocks of our flat connection and corresponding higher-genus polylogarithms will be modular tensors, and the resulting modular properties of the connection and the polylogarithms themselves
will be discussed in section \ref{sec:flcon} and \ref{sec:ghat}, respectively.

\subsubsection{Definition of modular tensors}
\label{sec:deftens}

Modular tensors are defined on Torelli space, which is the moduli space of compact Riemann surfaces endowed with a choice of canonical homology basis of $\mA$ and $\mB$ cycles. The significance of modular tensors has been articulated in the work of Kawazumi \cite{Kawazumi:lecture, Kawazumi:seminar, Kawazumi:paper} and  two of the authors of the present paper in \cite{DHoker:2020uid}. 
Modular tensors generalize modular forms at genus one by replacing the familiar automorphy factor $(C\tau+D)$ of $SL(2,\ZZ)$ discussed in section \ref{sec:ctaud} by an automorphy tensor $Q$ and its inverse $R= Q^{-1}$, 
\bea
\label{3.QR}
Q= Q(M, \Omega)  & = & C\Omega +D
\no \\
R= R(M, \Omega)  & = & (C\Omega +D)^{-1}
\eea
 for $M \in Sp(2h,\ZZ)$ and the matrices $C$ and $D$ given in (\ref{3.mod1}). The composition law for the automorphy tensors may be read off from the transformation properties of $\Omega$ given in (\ref{3.mod2}), 
 \bea
 Q(M_1M_2, \Omega) = Q\big ( M_1, (A_2\Omega+B_2)(C_2 \Omega +D_2)^{-1} \big ) Q(M_2, \Omega)
 \eea
In (\ref{3.mod2}) we already encountered the tensors $\bom_I$,  $\bom^I$, $Y_{I J} $, and its inverse $Y^{I J}$. In the notation (\ref{3.QR}) of the automorphy factors $Q$ and $R$, their transformation properties are given by,
\begin{align}
\tilde \bom_I & = \bom_{I'} R^{I'}{}_I
&
\tilde Y_{IJ} &= Y_{I' J'} \, \bar R^{I'}{}_I  \, R^{ J'}{}_J
\no \\ 
\tilde \bom^{ J} & = \bar Q^{ J} {}_{ J '} \, \bom ^{ J'} &
\tilde Y^{I J} & = Q^I{}_{I'} \, \bar Q^{ J }{}_{J'} \, Y^{I'  J'}
\label{ytrflaw}
\end{align} 
More generally, a modular tensor $\cT$ of arbitrary rank transforms as follows,
\bea
\tilde \cT^{I_1, \cdots, I_n;  J_1, \cdots,  J_{\bar n}} (\tilde \Omega) 
=  Q^{I_1}{}_{I_1'} \, \cdots \, Q^{I_n}{}_{I_n'} \, 
 \bar Q^{ J_1}{}_{ J_1'} \, \cdots \, \bar Q^{J_{\bar n} }{}_{J_{\bar n}' } \, 
 \cT^{I_1', \cdots, I_n'; J_1', \cdots, J_{\bar n}'} (\Omega)
 \label{tenstrf.01}
 \eea
The tensors $Y_{I J}$ and $Y^{I J}$ may be used to lower and raise indices, respectively,
and can be made to compensate any anti-holomorphic automorphy factor: instead of
the $\bar Q^{ J_i}{}_{ J_i'}$ in (\ref{tenstrf.01}), the tensor,
\bea
 \cU^{I_1, \cdots, I_n}_{ J_1, \cdots,  J_{\bar n}} (\Omega)
 = Y_{J_1 K_1} \cdots Y_{J_{\bar n} K_{\bar n}} \, \cT^{I_1, \cdots, I_n;  K_1, \cdots,  K_{\bar n}} (\Omega)
 \label{tenstrf.02}
\eea
exclusively transforms with holomorphic automorphy factors $Q^{I_i}{}_{I_i'}$ and $R^{J_i'}{}_{J_i}$,
\bea
\tilde \cU^{I_1, \cdots, I_n}_{ J_1, \cdots,  J_{\bar n}} (\tilde \Omega)
= Q^{I_1}{}_{I_1'} \, \cdots \, Q^{I_n}{}_{I_n'} 
R^{J_1'}{}_{J_1} \, \cdots \, R^{J_{\bar n}'}{}_{J_{\bar n}} \, 
 \cU^{I_1', \cdots, I_n'}_{ J_1', \cdots,  J_{\bar n}'} (\Omega)
 \label{tenstrf.03}
\eea
The above tensors may be reducible. Symmetrization, anti-symmetrization, and removal of the trace by contracting with factors of $Y_{IJ}$ or $\delta_I^J$ may be used to extract irreducible tensors. For genus $h$, the anti-symmmetrization of $h$ indices $I$ (resp.\ $J$ indices) produces a factor of $\det Q$ (resp.\ $\det \bar Q$).

\subsubsection{The interchange lemma}

The modular tensor defined by the following convolution,
\bea
\Phi^I{}_J(x) = \int _{\Sigma} d^2z \, {\cal G}(x,z) \, \bar \omega^I(z)   \omega_J(z) 
\label{polylog.28}
\eea
 was found to play a central role in the studies of more general modular tensors and their relations. In view of the second equation in (\ref{3.Arak}), the tensor $\Phi$ is trace-less, and therefore vanishes identically at genus one. In particular, it enters into the \textit{interchange lemma} which was stated and proved in \cite{DHoker:2020tcq, DHoker:2020uid},
\bea
\label{polylog.31}
\p_x \cG(x,y) \, \om_J(y)  + \p_y \cG(x,y) \, \om_J(x)  - \p_x \Phi^I{}_J (x) \, \om_I(y)  - \p_y \Phi^I{}_J (y) \, \om_I(x) =0
\eea
The genus-one version of this formula holds trivially by translation invariance on the torus and the vanishing of $\Phi$.  At higher genus, $\Phi$  may be viewed as compensating for the lack of translation invariance on higher-genus Riemann surfaces.

\subsubsection{Higher convolutions of the Arakelov Green function}
\label{sec:3.1}

The fundamental building blocks for a flat connection on higher-genus Riemann surfaces are modular tensors   
defined recursively by convolutions on $\Sigma$ as follows,
\begin{align}
\Phi ^{I_1 \cdots I_r}{}_J (x) & =  \int _{\Sigma}   d^2 z \, \cG(x,z) \, \bar \om^{I_1} (z) \, \p_z \Phi^{I_2 \cdots I_r}{}_J (z) \hskip -0.2in & r \geq 2
\label{polylog.29} \\
\cG ^{I_1 \cdots I_s} (x,y) & =  \int  _{\Sigma}  d^2 z \, \cG(x,z) \, \bar \om^{I_1} (z) \, \p_z \cG ^{I_2 \cdots I_s} (z,y) \hskip -0.2in & s \geq 1
\notag
\end{align}
where $\Phi^I{}_J(x)$ was defined  in (\ref{polylog.28}) and $\cG^{\emptyset}(x,y)=\cG(x,y)$. By construction, $\Phi ^{I_1 \cdots I_r}{}_J (x)$ and $\cG ^{I_1 \cdots I_s} (x,y)$ are  scalar functions of $x,y$ and $Sp(2h,\ZZ)$ tensors  of rank $r{+}1$ and $s$, respectively, with the purely holomorphic transformation law in (\ref{tenstrf.03}). The vanishing of the trace $\Phi ^{I_1 \cdots I_r}{}_{I_r} =0$ for arbitrary genus implies that $\Phi$-tensors for arbitrary $r\geq 1$ vanish identically for genus one. Furthermore, at genus one,  the derivatives of the tensor $ \cG^{I_1 \cdots I_s} $ for $I_1 = \cdots = I_s=1$ equal the Kronecker-Eisenstein integration kernels $f^{(s+1)}$ given in (\ref{polylog.04}) and (\ref{polylog.17}),
\beq
\partial_x \cG^{I_1 \cdots I_s} (x,y)\, \big|_{h=1} = - f^{(s+1)}(x{-}y| \tau)
\label{redGtens}
\eeq
From the differential relations satisfied by the Arakelov Green function in (\ref{3.Arak}) and (\ref{3.Arak2}) we derive  analogous differential relations satisfied by $\Phi ^{I_1 \cdots I_r}{}_J (x) $, 
\bea
\pbx \p_x \, \Phi^I{}_J(x) & = & - \pi \, \bar \omega^I(x) \, \omega_J(x) + \pi \, \delta^I_J \, \kappa(x)
\no \\
\pbx \p_x \,  \Phi ^{I_1 \cdots I_r}{}_J (x) & = &  - \pi \, \bar \om^{I_1} (x) \, \p_x \Phi ^{I_2 \cdots I_r}{}_J (x) 
\hskip 0.6in r \geq 2
 \label{polylog.33}
\eea
and  by $\cG ^{I_1 \cdots I_s} (x,y)$ for $s \geq 1$,
\bea
\label{polylog.34}  
\pbx \p_x \, \cG ^{I_1 \cdots I_s} (x,y) & = &  - \pi \, \bar \om^{I_1} (x) \, \p_x \cG ^{I_2 \cdots I_s} (x,y) 
\notag \\
\pby \p_x \, \cG ^{I_1 \cdots I_s} (x,y) & = &
 \pi  \, \p_x \cG ^{I_1 \cdots I_{s-1}}(x,y) \, \bar \om^{I_s} (y)   
 - \pi \, \partial_x \Phi^{I_1 \cdots I_s}{}_J(x) \, \bar \om^J(y) 
\eea
These relations will be fundamental in the construction of the flat connection at higher genus.

\subsubsection{Modular properties of convolutions}

The convolutions $\cG^{I_1 \cdots I_s}(x,y)$ and $\Phi^{I_1 \cdots I_r}{}_J(x)$ are modular tensors. Their transformation properties may be read off directly from those of $\bom_J$ and $\bar \bom^I$, 
\bea
\label{3.barom}
\tilde {\bar \bom} ^I(x) = Q^I{}_J \, \bar \bom^J(x)
\eea
see (\ref{ytrflaw}), and provide the following transformation rules, 
\bea
\tilde \cG ^{I_1 \cdots I_s} (x,y) & = & Q^{I_1} {}_{I_1'} \cdots Q^{I_s} {}_{I_s'} \cG ^{I_1' \cdots I_s'} (x,y) 
\no \\
\tilde \Phi ^{I_1 \cdots I_r}{}_J  (x) & = & Q^{I_1} {}_{I_1'} \cdots Q^{I_r} {}_{I_r'} \, 
\Phi^{I_1' \cdots I_r'} {}_{J'} (x) \, R^{J'}{}_J
\eea
We note that all automorphy tensors $Q,R$ in (\ref{3.QR}) are holomorphic on Torelli space.  Thus we may view $\cG^{I_1 \cdots I_s}(x,y)$ and $\Phi^{I_1 \cdots I_r}{}_J(x)$ as sections of holomorphic vector bundles over Torelli space, whose transition functions are given by the tensor $Q$ and its inverse $R$.

\subsection{Generating functions} 

For genus one, the functions $f^{(n)}(z|\tau)$ were obtained by expanding the Kronecker-Eisenstein series $\Omega(z,\alpha|\tau)$ in powers of the free parameter $\a$. At higher genus, we may also assemble the families of modular tensors, defined in the previous subsection, into generating functions.  To do so, we introduce a non-commutative  algebra freely generated by $B_I$ for $I=1,\cdots, h$, and that will soon be extended to a larger free algebra. 
We also fix an arbitrary auxiliary marked point $p$ on the Riemann surface $\Sigma$. With the help of the generators $B_I$, we introduce the following generating functions,
\begin{align}
\cH(x,p;B) &= \partial_x \cG(x,p) + \sum_{r=1}^\infty 
  \partial_x \cG^{I_1 I_2 \cdots I_r}(x,p) B_{I_1} B_{I_2} \cdots B_{I_r}
  \notag \\   
& = \partial_x \cG(x,p) + \partial_x \cG^{I_1}(x,p) B_{I_1}  
 + \partial_x \cG^{I_1I_2}(x,p) B_{I_1} B_{I_2} + \cdots 
 \notag \\
\cH_J(x;B)  &= \omega_J (x) + \sum_{r=1}^\infty 
  \partial_x \Phi^{I_1 I_2 \cdots I_r}{}_J(x) B_{I_1} B_{I_2} \cdots B_{I_r}
  \no \\
  &= \omega_J (x) + \partial_x \Phi^{I_1}{}_J(x) B_{I_1} 
+ \partial_x \Phi^{I_1 I_2}{}_{J} (x) B_{I_1} B_{I_2} + \cdots 
 \label{polylog.35}
\end{align}
Note that the modular tensors $\cG^{I_1 \cdots I_r}(x,p)$ and $\Phi^{I_1 \cdots I_r}{}_J(x)$ are not necessarily symmetric in their indices $I_1, \cdots, I_r$. This absence of symmetry is captured by the non-commutative nature of the algebra of the $B_I$. The differential relations (\ref{3.Arak}), (\ref{polylog.33}), and (\ref{polylog.34}) on the components imply the following differential relations on the generating functions, 
\begin{align}
\pbx \cH(x,p;B) & =   \pi \, \kappa (x) -  \pi \delta (x,p)   -  \pi \bar \om^I(x) B_I \, \cH(x,p;B)  
\notag \\
\partial_{\bar p} \cH(x,p;B) &=  \pi \delta (x,p)   + \pi \bar \om^I(p) 
\big(  \cH(x,p;B)  B_I -  \cH_I(x;B) \big)
\notag \\
\pbx \cH_J(x;B) & =  \pi \, \kappa (x) B_J  -  \pi \, \bar \om^I(x) B_I \, \cH_J (x;B)
 \label{polylog.36} 
\end{align}
By forming the combination,
\bea
\Psi_J(x,p;B) = \cH_J(x;B) - \cH(x,p;B)B_J
 \label{polylog.37} 
\eea
the differential form $\kappa$ in (\ref{polylog.24}) is found to cancel between the $\bar x$-derivative of both terms in (\ref{polylog.36}), and the result may be compactly written as follows,
\begin{align}
 \label{polylog.38}
\partial_{\bar x} \Psi_J (x,p;B) &=  \pi \delta(x,p) B_J  - \pi \bar \om^I(x) \, B_I \, \Psi_J (x,p;B) 
\no  \\
\partial_{\bar p} \Psi_J (x,p;B) &= - \pi \delta(x,p)  B_J + \pi \bar \om^I(p) \, \Psi_I(x,p;B) \,  B_J
\end{align}
The delta functions on the right-hand side signal the simple pole of  $\partial_x \cG(x,p) =  - \frac{1}{x-p}+{\cal O}(1)$ whereas all the tensorial integration kernels $\partial_x \Phi ^{I_1 \cdots I_r}{}_J  (x)$  and 
$\partial_x \cG ^{I_1 \cdots I_r} (x,p) $ with $r\geq 1$ are regular on the entire surface. This generalizes the 
pole structure of the $f^{(n)}$ to arbitrary genus where $f^{(1)}(x{-}p|\tau) =  \frac{1}{x{-}p}+{\cal O}(x{-}p)$ exhibits the only pole among the genus-one kernels.

\sm

To obtain tensorial modular transformations properties for the generating function (\ref{polylog.37}), the modular transformations of its components must be accompanied by the following transformation properties for the algebra generators $B_J$, 
\bea
\label{modother}
\tilde B_J & = &  B_{J'} R^{J'}{}_J 
\no \\
 \tilde \cH_J (x; \tilde B)  & = & \cH_{J'} (x;B) R^{J'}{}_J
\no \\
\tilde \Psi_J (x,p; \tilde B) & = & \Psi_{J'} (x,p;B) R^{J'}{}_J 
\eea
The generating function $\cH(x,p;B)$ is then invariant.

\subsection{The flat connection}
\label{sec:flcon} 

We are now ready to assemble all the results of the previous sections into a flat and modular invariant connection.

\sm

To do so, we begin by extending the algebra generated by the elements $B_I$ as follows. We introduce the Lie algebra $\cL$ freely generated by elements $a^I$ and $b_I$ for $I=1,\cdots, h$ and set $B_I = {\rm ad}_{b_I} = [b_I, \cdot]$. The algebra admits a dual grading counting independently the number of letters $a$ and the number of letters $b$ in each word, irrespective of the value of their indices. 
This algebra was already considered by Enriquez and Zerbini in their construction
of Maurer-Cartan elements in \cite{Enriquez:2021}, where the generators $a^I$ and $b_I$ correspond to generators of the fundamental group of the surface $\Sigma$ with the point $p$ removed. For a general reference on freely generated Lie algebras and their applications we refer to Reutenauer's book \cite{Reut}.

\sm

It remains to generalize the  term proportional to $(dz - d\bar z)b$ in the genus-one connection $\cJ_{\rm BL}$ in (\ref{polylog.06b}) to higher genus.  Since the single $b$ at genus one generalizes to a tensor $b_I$ at higher genus, it might seem natural to generalize $dz - d\bar z$ to the closed differential $\bom^I-\bar \bom^I$. Actually, this choice does not lead to a flat connection, but promoting the $\om^I$ part of this construction to $\cH^I$ in (\ref{polylog.35}) does the job. The result is the following theorem.

{\thm 
\label{3.thm1}
The connection $\cJ(x,p)$, on a Riemann surface $\Sigma$ of arbitrary genus $h$ with a marked point $p \in \Sigma$ and valued in the Lie algebra $\cL$  freely generated by the $2h$ elements $a^I, b_I$ with $I=1,\cdots,h$, is given in terms of $\bar \bom^I(x)$ and the generating functions $\cH^I(x;B)= Y^{IJ}\, \cH_J(x;B)$ and $\Psi_I(x,p;B)$ as follows, 
\bea
\cJ(x,p) = - \pi \, d \bar x \, \bar \om^I(x) \, b_I + \pi \, dx \, \cH^I (x;B) \, b_I + d x\, \Psi_I(x,p;B) \, a^I
 \label{polylog.39}  
 \eea
 where $B_I = {\rm ad}_{b_I} = [b_I, \cdot ]$. The connection $\cJ(x,p)$  is flat  and reproduces the Brown-Levin connection (\ref{polylog.06b}) at genus one.}
 
 \sm
 
To prove the theorem we begin by using the differential equations (\ref{polylog.36}) and (\ref{polylog.38}) satisfied by  the generating functions $ \cH^I (x;B)$ and $\Psi_I(x,p;B)$, and readily establish the following results, 
 \bea
 \label{polylog.41}
d_x \cJ(x,p) & = &
\pi d \bar x\wedge d x \Big \{ \delta(x,p)  [ b_I, a^I]   
- \pi  \bar \om ^I(x) \, B_I \, \cH^J (x;B) \, b_J 
 \\ && \hskip 0.8in
-  \bar \om^I(x)\, B_I \, \Psi_J(x,p;B) \, a^J \Big \}
\no \\
\cJ(x,p)  \wedge \cJ(x,p)  &=& 
 \pi d\bar x   \wedge d x  \Big ({-} \pi \, \bar \om ^I(x) \, B_I\, \cH^J(x;B) b_J 
 - \, \bar \om ^I(x) \, B_I \, \Psi_J (x,p;B) \, a^J \Big )
\no 
\eea
The difference of the two lines shows that the connection is flat away from $x=p$,
\bea
d_x\cJ(x,p) - \cJ(x,p) \wedge \cJ(x,p) = \pi d \bar x\wedge d x \,  \delta(x,p) \, [ b_I, a^I] 
\eea
To prove that the connection $\cJ(x,p)$ reduces to the non-holomorphic single-valued Brown-Levin connection, we specialize to the case of genus one and relabel $a^1=a$ and $b_1=b$. Since the tensor $\Phi^I{}_J$ of (\ref{polylog.28}) and its higher-rank versions of (\ref{polylog.29}) all vanish identically at genus one, the generating function $\cH^1(x;B) $ reduces to,
\bea
\cH^1(x;B) \,  \Big |_{h=1} = \om^1(x)= \frac{ \om_1(x)}{ \Im \tau }
\label{redhvector}
\eea
 so that the first two terms in (\ref{polylog.39}) combine to $\pi (dx-d\bar x) b/\Im \tau$ thereby reproducing the contributions $\sim (\Im \tau)^{-1}$ to the non-meromorphic Brown-Levin connection of (\ref{polylog.06b}). The third term in (\ref{polylog.39}) reproduces the Kronecker-Eisenstein series by (\ref{redhvector}) and (\ref{redGtens}),
 \bea
\Psi_1(x,p;B)  \,  \Big |_{h=1}  = \omega_1(x) - {\cal H}(x,p;B) B_1\, \Big |_{h=1}
= {\rm ad}_b \, \Omega(x{-}p,{\rm ad}_b|\tau)
 \eea
concluding the proof of Theorem \ref{3.thm1}.

\sm 

The expression (\ref{polylog.39}) for the flat connection is modular invariant for
suitable $Sp(2h,\ZZ)$ transformation rules of the generators $a^I,b_I$ to be stated
in the following theorem:
{\thm 
\label{thmmod}
Under a modular transformation $M \in Sp(2h,\ZZ)$, parametrized in (\ref{3.mod1}), which acts on $\bar \om^I$ as given in (\ref{3.barom}), on $B_I$, $\cH_I$, and $\Psi_I$ as given in (\ref{modother}), and on the Lie algebra generators $a^I$ and $b_I$ by, 
 \bea
 \label{3.thm1b}
a^I & \to & \tilde a^I = Q^I{}_J \, a^J + 2 \pi i \, C^{IJ} \, b_J
\no \\
b_I & \to & \tilde b_I = b_J \, R^J {}_I
\eea
the connection $\cJ(x,p)$ is invariant. In the  basis $(\hat a^I, b_I)$ of generators of the Lie algebra $\cL$,
\bea
\label{3.basis}
\hat a^I = a^I + \pi Y^{IJ} b_J
\eea
subject to
 \bea
 \label{3.atfon}
 \hat a ^I \to \tilde {\hat a}^I =  Q^I{}_J \, \hat a^J
 \eea
the connection $\cJ(x,p)$ takes on a simplified form,
\bea
\cJ(x,p) = - \pi \, d \bar x \, \bar \om^I(x) \, b_I  + d x\, \Psi_I(x,p;B) \, \hat a^I
 \label{polylog.39a}  
 \eea
 and is manifestly invariant under $Sp(2h,\ZZ)$.}
 
\sm

Proving modular invariance of $\cJ(x,p)$ is most transparently achieved by first carrying out the change of basis of (\ref{3.basis}) to the equivalent form (\ref{polylog.39a}) which immediately follows from the observation that $Y^{IJ} B_I b_J=0$ so that $\Psi_I(x,p;B) Y^{IJ} b_J = \cH^I(x;B) b_I$. The connection $\cJ(x,p)$, presented in the form of (\ref{polylog.39a}), is term-by-term invariant under the modular transformations of $\bar \om^I$,  $B_I$, $\Psi_I$, and $a^I,b_I$ stated in the theorem: in both of $d\bar x\, \bar \om^I(x)\, b_I$ and $dx \, \Psi_I(x,p;B) \, \hat a^I$, the respective ingredients transform with opposite automorphy factors as can be verified from (\ref{ytrflaw}), (\ref{modother}), (\ref{3.thm1b}) and (\ref{3.atfon}).
The modular invariance of $d\bar x\, \bar \om^I(x) \, b_I$ and $dx\, \Psi_I(x,p;B) \, \hat a^I$ established 
in this way completes the proof of Theorem \ref{thmmod}.

\sm

Finally, the connection $\cJ$ may be expanded in words with $r{+}1$ letters in the basis $(a^I,b_I)$,
\begin{align}
\label{3.Jsec}  
\cJ(x,p) & =  - \pi \, d\bar x \, \bar \om ^I(x) b_I + \pi \, dx\, \om^I(x) b_I 
+ \pi  \, dx \sum_{r=1}^\infty   \p_x \Phi^{I_1 \cdots I_r } {}_J(x)\, Y^{JK}\, B_{I_1} \cdots B_{I_r} \,  b_K 
\no \\ 
&\quad +\, dx \sum_{r=1}^\infty \Big (  \p_x \Phi^{I_1 \cdots I_r } {}_J(x)  - \p_x \cG^{I_1 \cdots I_{r-1}}  (x,p)  \delta_J^{I_r} \Big ) B_{I_1} \cdots B_{I_r} \,  a^J  
\end{align}
or, equivalently, in the basis $(\hat a^I, b_I)$,
\bea
\label{3.Jsec1}  
\cJ(x,p)  & =  &  \pi \Big( dx\,\om^I(x)  -  d\bar x\, \bar \om ^I(x) \Big) b_I 
\no \\ &&
+  dx \sum_{r=1}^\infty \! \Big (  \p_x \Phi^{I_1 \cdots I_r } {}_J(x)  -  \p_x \cG^{I_1 \cdots I_{r-1}}  (x,p)  \delta_J^{I_r} \Big ) B_{I_1}  \cdots  B_{I_r}   \hat a^J 
\eea
where we abbreviate $B_I = {\rm ad}_{b_I}$. These expressions will be useful in the subsequent section to illustrate the expansion of higher-genus polylogarithms, and the simplified representation in (\ref{3.Jsec1}) may be easily obtained from the change of  basis using (\ref{3.basis}).  

\sm

The higher-genus polylogarithms  obtained by expanding the path-ordered exponential (\ref{2.PO1}) of the connection $\cJ(x,p)$ in words of the generators $(a^I,b_I)$ or $(\hat a^I, b_I)$ will automatically be homotopy invariant. Examples to low  letter count will be given in the next section.

\newpage

\section{Higher-genus polylogarithms}
\label{sec:polylogs}
\setcounter{equation}{0}

In this section, the flat connection $\cJ$ assembled in the preceding section will be used to construct higher-genus polylogarithms. The flat connection $\cJ(x,p)$ of (\ref{polylog.39})  is well-defined and single-valued on two copies of $\Sigma$ but this property came at the cost of giving up meromorphicity in both of $x$ and $p$.  
We shall outline a method to restore meromorphicity in the first variable $x$ for certain combinations of integration kernels at the cost of giving up single-valuedness. Moreover, non-trivial evidence will be provided that the higher-genus polylogarithms in this section are closed under taking primitives. We close this section with a proposal for higher-genus analogues of elliptic associators.

\subsection{Construction of higher-genus polylogarithms}
\label{sec:hgenpoly}

The flat connection $\cJ(x,p)$ of Theorem \ref{3.thm1} integrates to a homotopy-invariant path-ordered exponential $\bGam(x,y;p)$, 
\begin{align}
\bGam(x,y;p) = \text{P} \exp  \int^x_y {\cal J}(t,p) 
 \label{polylog.51} 
 \end{align}
Expanding the path-ordered exponential in terms of the generators of the Lie algebra $\cL$ produces homotopy-invariant iterated integrals, as explained in the preamble of section \ref{sec:one}, 
\begin{align}
\bGam(x,y;p) = 1+ \sum _{\mw} \mw \, \Gamma (\mw;x,y;p)
 \label{polylog.51a} 
 \end{align}
 Here, the sum over $\mw$ is over all words, containing at least one letter, made out of the alphabet of the Lie algebra generators introduced in section \ref{sec:flcon}. In this way, the path-ordered exponential is the generating function for the iterated integrals $\Gamma (\mw;x,y;p)$ and, for each word $\mw$, defines a polylogarithm $\Gamma (\mw;x,y;p)$. Their iterated-integral definition via (\ref{polylog.51}) and (\ref{polylog.51a})  implies the following shuffle property,
\bea
 \Gamma (\mw_1;x,y;p) \cdot \Gamma (\mw_2;x,y;p) =  \sum_{ \mw \in \mw_1 \shuffle \mw_2 } \Gamma (\mw;x,y;p) 
 \label{shuffprop}
 \eea
 which holds in identical form for  the polylogarithms (\ref{polylog.00}) and (\ref{polylog.08}) at genus zero and genus one. Following the standard shuffle product, the sum over $\mw \in \mw_1 \shuffle \mw_2$  includes all ordered sets obtained from $\mw_1 \cup \mw_2$ that preserve the order within the individual words $\mw_1,\mw_2$.

 \sm
 
In Theorem \ref{3.thm1} and Theorem \ref{thmmod}, the modular invariant connection $\cJ(x,p)$ was expressed in terms of two different bases for the same Lie algebra $\cL$ in which $\cJ(x,p)$ takes its values. In the first  basis $(a^I,b_I)$ the relation with the Brown-Levin connection at genus one is manifest,  while the second basis $(\hat a^I, b_I)$ leads to polylogarithms $ \Gamma (\mw;x,y;p)$ that transform as modular tensors by the $Sp(2h,\ZZ)$-invariance of $\cJ(x,p)$ established in Theorem \ref{thmmod}. The expansion in either basis will be of interest, and both will be pursued in the sequel.

\sm

In order to avoid cluttering, we compactly label the polylogarithms in the expansion of (\ref{polylog.51a}) through an ordered sequence of upper and lower indices encoding the accompanying words in either the basis $(a^I,b_I)$ or the basis $(\hat a^I, b_I)$,
\bea
\Gamma_{\ldots I\ldots }{}^{\ldots J\ldots}(x,y;p) & = & \Gamma(\ldots a^I\ldots b_J \ldots;x,y;p)
\no \\
\hat \Gamma_{\ldots I\ldots }{}^{\ldots J\ldots}(x,y;p) & = & \Gamma(\ldots \hat a^I\ldots b_J \ldots;x,y;p)
 \label{polylog.53}
\eea
For example, for words with at most two letters in the basis $(a^I,b_I)$, the expansion (\ref{polylog.51a}) takes the following form, 
\begin{align}
\bGam(x,y;p) & =   1 + a^I \Gamma_I(x,y;p) + b_I \Gamma^I(x,y;p) 
+ a^I a^J \Gamma_{IJ}(x,y;p)
\no \\ 
&\quad  
+ b_I b_J \Gamma^{IJ}(x,y;p)  + a^I b_J \Gamma_{I}{}^{J}(x,y;p)+ b_I a^J \Gamma^{I}{}_{J}(x,y;p)+ \cdots 
\label{4.gam}
\end{align}
while in the basis $(\hat a^I, b_I)$ it is given by,
\begin{align}
\bGam(x,y;p) & =   1 + \hat a^I \hat \Gamma_I(x,y;p) + b_I \hat \Gamma^I(x,y;p) 
+ \hat a^I \hat a^J \hat \Gamma_{IJ}(x,y;p)
\no \\ 
&\quad
+ b_I b_J \hat \Gamma^{IJ}(x,y;p)  
+ \hat a^I b_J \hat \Gamma_{I}{}^{J}(x,y;p)
+ b_I \hat a^J \hat \Gamma^{I}{}_{J}(x,y;p)+ \cdots 
\label{4.gama}
\end{align}
Identifying term by term in both expansions gives the relations $\Gamma _I = \hat \Gamma _I$ and $\Gamma _{IJ} = \hat \Gamma _{IJ}$ to be established in all generality in (\ref{simp.01}), as well as the following relations to this order,
\bea
\hat \Gamma ^I \, & = &  \Gamma ^I  - \pi Y^{IJ} \,  \Gamma _J
\no \\
 \hat\Gamma ^I {}_J & = & \Gamma ^I{}_J   - \pi Y^{IK} \,  \Gamma _{KJ}
\no \\
 \hat \Gamma _I {}^J & = &\Gamma _I{}^J   - \pi \,   \Gamma_{IK} \, Y^{KJ}
\no \\
 \hat  \Gamma ^{IJ} & = &\Gamma ^{IJ}    - \pi Y^{IK} \,  \Gamma _K {}^J - \pi  \,   \Gamma ^I{}_K \, Y^{KJ}
+ \pi^2 \, Y^{IK}  \,   \Gamma _{KL} \, Y^{LJ}
\label{convab}
\eea
We have suppressed the common arguments $(x,y;p)$ of the polylogarithms in order to avoid unnecessary clutter.

\subsubsection{Tangential end-point regularization and specialization to genus one}

Apart from their dependence on the endpoints $x,y$ of the integration path in (\ref{polylog.51}) and on  the moduli of $\Sigma$,  the higher-genus polylogarithms defined in  (\ref{polylog.51}) also depend on the marked  point $p$ that enters the connection ${\cal J}(t,p)$ in (\ref{polylog.39}). Setting $p=x$ or $p=y$ leads to endpoint divergences caused by the simple pole of $\partial_t {\cal G}(t,p) = - \frac{1}{t-p} + {\cal O}(1)$ which we shall shuffle regularize using the procedure introduced  in \cite{Broedel:2014vla}. This may be done by shifting  one or the other of the endpoints in the exponent of (\ref{polylog.51}) by $|\ep| \ll 1$ along a prescribed tangent vector \cite{Deligne:1989, Brown:2014, Panzer:2015ida, Abreu:2022mfk},
\bea
\int^{x-\ep}_y {\cal J}(t,p) \ {\rm if} \ p=x \ \qquad {\rm or} \qquad  \int^x_{y+\ep} {\cal J}(t,p) \ {\rm if} \ p=y
 \label{polylog.54}
\eea
expanding the regularized integral in powers of $\ep$, and defining the  value of the integral as the  term of order zero in the expansion, thus omitting divergent terms such as $\ln ( 2 \pi i \ep)$.  

\sm 

At genus one, the regularization procedure of \cite{Broedel:2014vla} leads to the elliptic polylogarithms (\ref{polylog.08}) of Brown-Levin which are obtained by setting $y=p=0$ in  (\ref{polylog.51}) or (\ref{polylog.53}), 
\bea
\Gamma(\ldots a \ldots b \ldots ;x) = \Gamma_{\ldots 1\ldots}{}^{\ldots 1 \ldots}(x,0;0 ) \, \big|_{h=1}
 \label{polylog.55}
\eea

\subsection{The special case of polylogarithms for words without $b_I$}

The polylogarithms associated with words $\mw$ that do not involve any of the letters $b_I$ are the same in the bases $(a^I,b_I)$ and $(\hat a^I, b_I)$ and are given by the following simple formula, 
\bea
\Gamma_{I_1 I_2\cdots I_r}(x,y;p) 
= \hat \Gamma_{I_1 I_2\cdots I_r}(x,y;p) 
= \int^x_y \bom_{I_1}(t_1) \int^{t_1}_y  \bom_{I_2}(t_2) \cdots  \int^{t_{r-1}}_y \bom_{I_r}(t_r)  
\label{simp.01}
\eea
which makes clear that these polylogarithms are actually independent of the marked point $p$ and confirms explicitly that they are modular tensors of rank $r$. For the case $r=1$, we recover the basic Abelian integrals. For $r=2$, a particular combination of the $\mA$-cycle monodromy in $x$ gives the Riemann vector \cite{Fay:1973} with base point $z_0$, 
\bea
\Delta _I(z_0) = - \half - \half \Omega_{II} + \sum_{J\not= I} \oint _{\mA_J}  \bom_J(z) \int ^z _{z_0} \bom_I
\eea
For $r \geq 2$, the simple subclass (\ref{simp.01}) of polylogarithms generalizes Abelian integrals to modular tensors of rank $r$. They obey the differential equations, 
\bea
\partial_x \Gamma_{I_1 I_2\cdots I_r}(x,y;p) = \om_{I_1}(x) \Gamma_{ I_2\cdots I_r}(x,y;p)
\label{simp.02}
\eea
and the simplest instance of the shuffle property (\ref{shuffprop}) reads,
\bea
\Gamma_{I}(x,y;p)\cdot \Gamma_{J}(x,y;p) = \Gamma_{IJ}(x,y;p) + \Gamma_{JI}(x,y;p)
\eea
Specializing to genus $h=1$ they may be evaluated explicitly,
\bea
\Gamma_{\underbrace{11\cdots 1}_r}(x,y;z) \, \big|_{h=1}= \frac{1}{r!} \, (x{-}y)^r
\label{simp.03}
\eea

\subsection{Low letter count polylogarithms in the basis $(a^I,b_I)$} 

Using the expansion (\ref{4.gam}) of the path-ordered exponential $\bGam(x,y;p)$ in powers of the connection combined with the expansion of the flat connection $\cJ(x,p)$ in (\ref{3.Jsec}),
\begin{align}
\label{expJ3}
\cJ(x,p) &=  
\bom_I a^I + \pi (\bom^I - \bar \bom^I  ) b_I 
+ dx \, \bigg( \partial_x \Phi^I{}_J(x)[b_I, a^J] - \partial_x \cG(x,p) [b_I,a^I] 
\notag \\
&\hskip 0.55in
+ \pi \partial_x \Phi^{I}{}_J(x) Y^{JK} [ b_I,b_K] 
+ \partial_x \Phi^{IJ}{}_K(x) \big[b_I, [ b_J,a^K] \big]
 \notag \\
& \hskip 0.55in
 - \partial_x \cG^I(x,p) \big[ b_I, [b_J,a^J] \big]  
 + \pi \partial_x \Phi^{IJ}{}_K(x) Y^{KL} 
 \big[ b_I, [b_J,b_L]\big] 
+ \cdots \bigg)
\end{align}
in terms of words in the alphabet $a^I, b_I$, we construct  polylogarithms for words that contain the letters $b_I$ as well as $a^I$.  For the single-letter word $b_I$, we obtain, 
\begin{align}
 \label{polylog.56} 
 \Gamma^I(x,y;p) &= \pi \int^x_y ( \bom^I  - \bar  \bom^I )
\end{align}
which is independent of $p$ but, as expected in the basis $(a^I, b_I)$, not a modular tensor.
For double-letter words with at least one letter $b_I$, we obtain, 
\begin{align}
 \Gamma^{IJ} (x,y;p) &= \pi   \int_y^x 
\bigg(
dt \, \big(\partial_t \Phi^I{}_K(t) Y^{KJ} - \partial_t \Phi^J{}_K(t) Y^{KI} \big)
+ \pi \big( \bom^I(t)  - \bar  \bom^I(t) \big) \int^t_y ( \bom^J  - \bar  \bom^J )
\bigg)  \no \\
\Gamma^J{}_I (x,y;p) &= \int_y^x \bigg( dt \, \partial_t \Phi^J{}_I(t) - dt \, \partial_t {\cal G}(t,p) \delta^J_I
 + \pi \big( \bom^J(t)  - \bar  \bom^J(t) \big)
   \int^t_y \bom_I  \bigg)\no \\
\Gamma_I{}^J (x,y;p) &= \int_y^x \bigg( {-} dt \, \partial_t \Phi^J{}_I(t) + dt \, \partial_t {\cal G}(t,p) \delta^J_I
 + \pi \bom_I (t)  \int^t_y  ( \bom^J  - \bar  \bom^J )  \bigg)  
 \label{polylog.99} 
\end{align}
The entry $\Gamma^{IJ}$ and the off-diagonal components of $\Gamma^J{}_I$ and $\Gamma _I{}^J$ are  independent on $p$.

\subsubsection{Simplified representations}

The polylogarithms in (\ref{polylog.56}) and (\ref{polylog.99}) with upper indices admit simplified representations in terms of (\ref{simp.01}), their complex conjugates and contractions with $Y^{IJ}$. For words with a single letter $b_I$ we have,
\bea
\label{simp.04} 
\Gamma^{I}(x,y;p) &= \pi Y^{IJ} \big( \Gamma_{J}(x,y;p) - \overline{ \Gamma_{J}(x,y;p) } \big)
\eea
while for two-letter words that contain at least one $b_I$, we have, 
\begin{align}
\label{simp.04a} 
\Gamma_I{}^{J}(x,y;p) &= \pi Y^{JK} \Gamma_{IK}(x,y;p)
+  \int^x_y dt \, \Big( 
{-} \partial_t \Phi^J{}_I(t) + \delta^J_I \partial_t \cG(t,p)
- \pi \omega_I(t) Y^{JK} \overline{ \Gamma_K(t,y;p) }
\Big) 
 \notag \\
\Gamma^I{}_J(x,y;p) &=  \pi Y^{IK} \big( \Gamma_{KJ}(x,y;p)
-  \Gamma_{J}(x,y;p) \overline{ \Gamma_{K}(x,y;p) }
\big) \notag \\
&\quad
+  \int^x_y dt \, \Big( 
 \partial_t \Phi^I{}_J(t) - \delta^I_J \partial_t \cG(t,p)
+ \pi \omega_J(t) Y^{IK} \overline{ \Gamma_K(t,y;p) }
\Big) 
\notag \\
\Gamma^{ IJ }(x,y;p)&= \pi^2 Y^{IK} Y^{JL}
\Big(
 \Gamma_{KL}(x,y;p)
 +\overline{  \Gamma_{KL}(x,y;p) }
 -  \overline{ \Gamma_{K}(x,y;p) }  \Gamma_{L}(x,y;p)
\Big) \notag \\
&\quad + \pi \int^x_y dt \, 
\Big(
\partial_t \Phi^{I}{}_K(t) Y^{KJ} - \partial_t \Phi^{J}{}_K(t) Y^{KI} \notag \\
&\quad \quad \quad
+ \pi \omega^J(t) Y^{IK}  \overline{ \Gamma_{K}(t,y;p) } 
-  \pi \omega^I(t) Y^{JK}  \overline{ \Gamma_{K}(t,y;p) } 
\Big) 
\end{align}
These expressions already illustrate the general fact that the indices on these polylogarithms cannot be simply raised or lowered by contraction with $Y_{IJ}$ or its inverse. Indeed,  $\Gamma^{I}(x,y;p) $ in (\ref{simp.04}) is not obtained by contracting $ \Gamma_{J}(x,y;p) $ with $Y^{IJ}$. 

\sm

Homotopy-invariance of the higher-genus polylogarithms in (\ref{simp.04})  and (\ref{simp.04a}) can be seen directly from the fact that the integrands  with respect to $t$ entering $\Gamma_I{}^{J}(x,y;p),\Gamma^I{}_{J}(x,y;p)$ 
and $\Gamma^{ IJ }(x,y;p)$ are meromorphic in $t$. The vanishing of the respective
$\partial_{\bar t}$-derivatives can be checked via (\ref{3.Arak}), (\ref{polylog.33}) and (\ref{simp.02}).
The shuffle relations $\Gamma^{ I }\cdot \Gamma^{ J }=\Gamma^{ IJ }+\Gamma^{ JI }$ and
$\Gamma^{ I }\cdot \Gamma_{ J }= \Gamma^I{}_J + \Gamma_J{}^I$ in (\ref{shuffprop})
are easily verified from the expressions in (\ref{simp.04}) and (\ref{simp.04a}).

\subsection{Low letter count polylogarithms in the basis $(\hat a^I,b_I)$} 
\label{sec:ghat}

The polylogarithms $\hat \Gamma (x,y;p)$ in the basis $(\hat a^I,b_I)$ defined by the expansion (\ref{4.gama}) are 
modular tensors by the $Sp(2h,\mathbb Z)$ invariance of the connection ${\cal J}(x,p)$
established in Theorem \ref{thmmod}. For words involving only $\hat a^I$ letters and no $b_I$ letters, the expressions were already given in (\ref{simp.01}). For the general case, it will be convenient to introduce the following expansion of the generating function $\Psi_I(x,p;B)$,
\bea
\Psi _J (x,p;B) & = & \om_J(x) + \sum_{r=1}^\infty B_{I_1} \cdots B_{I_r} \, f^{I_1 \cdots I_r} {}_J  (x,p)
\no \\
f^{I_1 \cdots I_r} {}_J (x,p) & = & \p_x \Phi ^{I_1 \cdots I_r} {}_J(x) - \p_x \cG^{I_1 \cdots I_{r-1}} (x,p) \delta ^{I_r}_J
\eea 
Expanding the connection $\cJ(x,p)$ accordingly, 
\bea
\cJ(x,p) = - \pi \bar \bom^I(x) b_I + \Big ( \bom_I(x)  \hat a^I + f^J {}_I(x,p) [b_J, \hat a^I] + f^{JK} {}_I(x,p) [b_J, [b_K, \hat a^I] ] + \cdots \Big )
\quad
\eea
we compute the polylogarithms for a one- and two-letter words, starting with the coefficient of the letter $b_I$, 
\bea
\hat \Gamma ^I(x,y;p) = - \pi \int^x _y \bar \bom^I
= - \pi Y^{IK} \,  \overline{ \Gamma_K (x,y;p) } 
\label{hatex}
\eea
This example illustrates that, also in the $(\hat a^I,b_I)$ basis, indices of polylogarithms 
$\hat \Gamma$ cannot be raised or lowered via $Y^{IJ}$ or $Y_{IJ}$.
For two-letter words that contain at least one letter $b_I$,
the simplest examples (\ref{simp.04a}) translate into,
\begin{align}
\hat \Gamma ^{IJ} (x,y;p) & =  \pi^2 \int _y^x \bar \bom^I(t_1) \int ^{t_1} _y \bar \bom^J
= \pi^2 Y^{IK} Y^{JL} \, \overline{ \Gamma_{KL}(x,y;p) } 
\no \\
\hat \Gamma _I {} ^J  (x,y;p) & =  - \int _y^x dt \, 
\bigg( f^J{}_I(t,p) + \pi \, \omega_I(t) \int^t_y \bar \bom^J \bigg) 
\no \\
\hat \Gamma ^I{}_J (x,y;p) & =   \int _y^x dt \,
\bigg( f^I{}_J(t,p) + \pi\, \omega_J(t) \int^t_y  \bar \bom^I \bigg)
- \pi Y^{IK} \, \overline{ \Gamma_K (x,y;p) }  \, \Gamma_J(x,y;p) 
\label{hatex.02}
\end{align}
under the conversion (\ref{convab}) between the $(a^I,b_I)$ and $(\hat a^I,b_I)$ basis.
These examples line up with the general transformation law of higher-genus polylogarithms $\hat\Gamma$ in the expansion (\ref{4.gama}) as modular tensors with holomorphic automorphy factors $R^{I'}{}_I $ and $Q^J{}_{J'} $ in (\ref{3.QR}),
\bea
\tilde{\hat \Gamma}_{\cdots I \cdots}{}^{\cdots J \cdots}(x,y;p) =
\cdots R^{I'}{}_I  \cdots Q^J{}_{J'} \cdots \hat \Gamma_{\cdots I' \cdots}{}^{\cdots J' \cdots}(x,y;p)
\label{trfghat}
\eea

\subsubsection{Genus-one illustration of the modular properties}

We shall now illustrate how modularity is realized in the last two polylogarithms in equation (\ref{hatex.02}) specialized to genus one. Upon specializing the expansion (\ref{4.gama}) 
and transformation law (\ref{trfghat}) to genus one,
any elliptic polylogarithm $\hat \Gamma_{\ldots I\ldots }{}^{\ldots J\ldots}(x,y;p) |_{h=1}$ 
with $m$ uppercase indices and $n$ lowercase
indices transforms as a modular form of $SL(2,\mathbb Z)$ with weight $(m{-}n,0)$.
In particular, the genus-one incarnation of the second example in (\ref{hatex.02}),
\bea
\! \! \! \! \! \! \! \hat \Gamma _I {} ^J (x,y;p) \, \big|_{h=1} = 
\tilde \Gamma\big( \smallmatrix 1 \\ p \endsmallmatrix;y|\tau \big) 
- \tilde \Gamma\big( \smallmatrix 1 \\ p \endsmallmatrix;x|\tau \big)
+ \frac{\pi }{\Im \tau } \, \big[
\tfrac{1}{2}(y{-}p)^2 -\tfrac{1}{2}(x{-}p)^2 
+ (\bar y {-}\bar p) (x{-}y) \big]
\eea
is modular invariant by this counting. Indeed, $SL(2,\mathbb Z)$-invariance
of the term $\sim  \frac{(\bar y {-}\bar p) (x{-}y)}{\Im \tau}$ is manifest whereas
modular invariance of the leftover expression $\sim 
\frac{(y{-}p)^2}{2\Im \tau} -\frac{(x{-}p)^2}{2\Im \tau}$ and
\bea
\tilde \Gamma\big( \smallmatrix 1 \\ p \endsmallmatrix;y|\tau \big) 
- \tilde \Gamma\big( \smallmatrix 1 \\ p \endsmallmatrix;x|\tau \big)
=  \ln \tet(y{-}p| \tau) - \ln \tet(x{-}p| \tau)
\eea
relies on cancellations between the respective $SL(2,\mathbb Z)$-transformations
obtained from (\ref{tettrfm}).

\sm

For the endpoint divergences of higher-genus polylogarithms $\hat \Gamma_{\cdots I \cdots}{}^{\cdots J \cdots}(x,y;p)$ at $p=x$ or $p=y$, the regularization prescription (\ref{polylog.54}) necessitates the specification of a tangent vector. It remains to be investigated which choices of tangent vectors preserve the tensorial modular transformation of
$\hat \Gamma_{\cdots I \cdots}{}^{\cdots J \cdots}(x,y;p)$ at $p\in \{x,y\}$. 

\subsection{Meromorphic variants}

The first example of the higher-genus polylogarithms in (\ref{simp.04a}) can be written as,
\begin{align}
\Gamma_I{}^{J}(x,y;p) &=
\Gamma(a^I b_J;x,y;p)
\label{merosec.01} \\
&= \int^x_y dt \, \Big( 
{-} \partial_t \Phi^J{}_I(t) + \delta^J_I \partial_t \cG(t,p)
+ \pi \omega_I(t) Y^{JK} \big(  \Gamma_{K}(t,y;p)  - \overline{ \Gamma_K(t,y;p) }
\Big)\notag
\end{align}
Upon specializing to genus $h=1$ and setting $p=y=0$, this reproduces the Brown-Levin polylogarithm $\Gamma(ab;p|\tau)  = - \tilde \Gamma\big( \smallmatrix 1 \\ 0 \endsmallmatrix;p|\tau \big )$ in (\ref{polylog.12a}),
 namely the integral over the meromorphic kernel $g^{(1)}(t|\tau)$ in (\ref{exft01}).
 Accordingly, one may view the integrand with respect to $t$ in the second line of (\ref{merosec.01})
 as a higher-genus uplift of the  Kronecker-Eisenstein kernel $g^{(1)}(t|\tau)$,
 \bea
g^J{}_I(t,y;p) =  \partial_t \Phi^J{}_I(t) - \delta^J_I \partial_t \cG(t,p)
- 2 \pi i \omega_I(t) Y^{JK}  \Im \int^t_y \bom_K 
\label{merosec.02}
 \eea
Indeed, the Laplace equations (\ref{3.Arak}) and (\ref{polylog.33}) of $\cG(t,p)$ and $\Phi^J{}_I(t)$
readily imply meromorphicity in $t$,
\bea
\partial_{\bar t} g^J{}_I(t,y;p) = \pi \delta^J_I \delta(t,p)
\label{merosec.03}
\eea
verifying homotopy invariance of (\ref{merosec.01}). However, (\ref{merosec.02}) is not meromorphic in the endpoint~$y$ of the integration path or the second argument $p$ of the flat connection. At genus $h=1$, setting $y\rightarrow p$ readily reduces $g^J{}_I(t,y;p) $ to  the kernel $g^{(1)}(t{-}y|\tau)$ meromorphic in both $t$ and $y$. Starting from genus $h\geq 2$, by contrast, (\ref{merosec.02}) at $y=p$ does not yield a meromorphic
function of two points $t,y$ on $\Sigma$ since
\bea
\partial_{\bar y} \big( g^J{}_I(t,y;p) \, \big|_{y=p} \big)= - \pi \delta^J_I \delta(t,y)
+ \pi \Big( \delta^J_I \omega_K(t) \bar \omega^K(y) - \omega_I(t) \bar \omega^J(y) \Big)
\label{merosec.11}
\eea
Corrections of $g^J{}_I(t,y;p) \, \big|_{y=p} $ by abelian integrals $  \omega_I(t)   \Im \int^t_y \bom^J $ or $ \delta^J_I  \omega_K(t)   \Im \int^t_y \bom^K$ do not suffice to attain simultaneous meromorphicity in $t$ and $y$. Still, one
can add combinations of abelian integrals  $  \omega_I(t)   \Im \int^y_p \bom^J $ and $\delta_I^J   \omega_K(t)   \Im \int^y_p \bom^K$ to render $g^J{}_I(t,y;p)$ meromorphic in both $t$ and $y$ at the cost of a separate dependence on a third marked point $p$.

\sm

One can similarly take the Brown-Levin polylogarithms $\Gamma(a b\cdots b;p|\tau) 
  = (-1)^n \tilde \Gamma\big( \smallmatrix n \\ 0 \endsmallmatrix;p|\tau \big)$ in (\ref{polylog.12a}) 
with $n\geq 2$ letters $b$ as a starting point to motivate higher-genus analogues of
the meromorphic kernels $g^{(n\geq 2)}$ in (\ref{polylog.10}), e.g.
\begin{align}
\Gamma_I &{}^{JK}(x,y;p) =
\Gamma(a^I b_J b_K;x,y;p)
\label{merosec.04} \\
&= \int^x_y dt \, \bigg\{
\partial_t \Phi^{KJ}{}_I(t) - \delta^J_I \partial_t {\cal G}^K(t,p) - \pi \partial_t \Phi^J{}_I(t)
Y^{KL}  \big(  \Gamma_{L}(t,y;p)  - \overline{ \Gamma_L(t,y;p) }
\big) \notag \\
&\quad \quad \quad + \pi  \delta_I^J \partial_t {\cal G}(t,z) Y^{KL}  \big(  \Gamma_{L}(t,y;p)  - \overline{ \Gamma_L(t,y;p) } \big) \notag \\
&\quad \quad\quad 
+ \pi^2 \omega_I(t) Y^{JL} Y^{KM} \big(
\Gamma_{LM}(t,y;p) 
-  \overline{ \Gamma_L(t,y;z) } \Gamma_{M}(t,y;p) 
+  \overline{ \Gamma_{LM}(t,y;p) }
\big)\notag \\
&\quad \quad\quad  + \pi \omega_I(t) \int^t_y dt' \big(   \partial_{t'} \Phi^J{}_L(t') Y^{LK}
 - \pi \omega^J(t') Y^{KL}  \overline{ \Gamma_L(t',y;p) } 
-(J\leftrightarrow K)\big)
\bigg\}
\notag
\end{align}
Again, the integrand on the right-hand side is meromorphic in $t$ and may be viewed as a higher-genus generalization $g^{KJ}{}_I$ of $g^{(2)}$. However, simultaneous meromorphicity in $t$ and $y$ cannot be attained without admitting 
dependences on additional marked points $z$.
Hence, our construction does not suggest any straightforward generalizations  of the meromorphic Kronecker-Eisenstein coefficients $g^{(n)}(t{-}y|\tau)$ to higher  genus which meromorphically depend on two points $t,y$ on the surface without any reference to additional points. Instead, the tensors $\partial_t \Phi^{I_1\ldots I_r}{}_J(t)$ and $\partial_t {\cal G}^{I_1\ldots I_s}(t,y)$  of section \ref{sec:gtensor} naturally generalize the doubly-periodic
Kronecker-Eisenstein kernels $f^{(n)}(t{-}y|\tau)$ to arbitrary genus. 

\sm

The above examples motivate the study of gauge transformations $U(x,p)$, 
whose action on the connection is given by,
\bea
\tilde \cJ(x,p) =  U(x,p)  {\cal J}(x,p)  U(x,p)^{-1} + \big(d U(x,p)\big) U(x,p)^{-1}
\eea
and which induce the following transformation on the path-ordered exponential,
\bea
\tilde \bGam(x,y;p)=U(x,p)  \bGam(x,y;p) U(y,p)^{-1}
\eea 
such that $\tilde \bGam(x,y;p)$ is meromorphic in $x$ and $y$ to all orders in $a^I$ and $b_J$. 
Based on the generalized abelian integrals (\ref{simp.01}) and their complex conjugates, 
it is not difficult to construct a gauge transformation that implements meromorphicity 
in $x$ and yields generating series of the elliptic polylogarithms (\ref{polylog.11}) upon 
specialization to genus one. Refined choices of $ U(x,p) $ that additionally preserve 
the vanishing of $\mA$-cycle monodromies of ${\cal J}(x,p)$ in $x$ and make contact
with the meromorphic connections in the work of Enriquez \cite{Enriquez:2011} 
and Enriquez-Zerbini \cite{Enriquez:2021,Enriquez:2022} are currently
under investigation.

\subsection{Closure under taking primitives} 
\label{sec:close}

The closure of elliptic polylogarithms under taking primitives crucially hinges on translation invariance at genus one and the Fay identity among Kronecker-Eisenstein kernels \cite{BrownLevin, Broedel:2014vla, Broedel:2017kkb}. We shall now present evidence that a similar closure holds for products of the higher-genus polylogarithms in (\ref{polylog.51}) with the integration kernels in the connection (\ref{polylog.39}). As we will see below, the two key mechanisms for closure under taking primitives are the  interchange lemma (\ref{polylog.31}) and Fay identities at higher genus.

\subsubsection{Implications of the interchange lemma}

Given that the path-ordered exponential in (\ref{polylog.51}) only involves integrals 
over the first argument $t$ of the connection ${\cal J}(t,p)$ in the integrand, the primitive of
$\partial_t {\cal G}(t,p) \omega_I(p)$ with respect to $p$ is not obvious from the definition
of higher-genus polylogarithms. However, the interchange lemma (\ref{polylog.31}) yields an alternative
representation of $\partial_t {\cal G}(t,p) \omega_I(p)$ where its primitive  with respect to $p$ can be readily found in terms of higher-genus polylogarithms via (\ref{simp.02}) and (\ref{simp.04}),
\begin{align}
\partial_t {\cal G}(t,p) \omega_J(p)  
&=  \partial_p \Phi^I{}_J(p) \omega_I(t) + \partial_t \Phi^I{}_J(t) \omega_I(p)
-\partial_p {\cal G}(p,t) \omega_J(t)  \label{polylog.61} \\
&= \partial_p \Big({-} \Gamma_J{}^I(p,t;t) \omega_I(t) + \Gamma_I(p,t;t) \partial_t \Phi^I{}_J(t) \notag \\
&\quad + \pi \omega_I(t) Y^{IK} \big[ \Gamma_{JK}(p,t;t) 
-  \Gamma_{J}(p,t;t)  \overline{  \Gamma_{K}(p,t;t) } \big] \Big)
\notag
\end{align}
Similarly, convolutions of the interchange lemma (\ref{polylog.31}) with Arakelov Green functions
allow to rewrite higher-rank expressions $\partial_t {\cal G}^{I_1 I_2\cdots I_s}(t,p) \omega_J(p)$ as
(see section 5 of \cite{DHoker:2024ozn} for a proof),
\begin{align}
\partial_t {\cal G}^{J_1 J_2 \cdots J_s}(t,p) \omega_I(p) &= (-1)^{s-1} \partial_p {\cal G}^{J_s\cdots  J_2 J_1}(p,t) \omega_I(t)   \label{ik.13}   \\
&\quad +\bigg\{ 
 \sum_{i=1}^s   (-1)^i  \partial_t \Phi^{J_1 J_2 \cdots J_{s-i} K}{}_I(t)
\partial_p \Phi^{J_s J_{s-1} \ldots J_{s-i+1}}{}_K(p) \notag \\
&\quad \quad \quad 
+ \partial_t \Phi^{J_1 J_2 \cdots J_s K}{}_I(t) \omega_K(p) 
+ (-1)^s \Big(  \smallmatrix t \leftrightarrow p \\
J_1 J_2 \cdots J_s \leftrightarrow J_s \cdots  J_2 J_1 \endsmallmatrix \Big) \bigg\}
\notag
\end{align}
where the primitives of the right-hand side with respect to $p$ are accessible from
the higher-genus polylogarithms in (\ref{polylog.51}) and their complex conjugates.
The instruction to add $(-1)^s$ times the image under the simultaneous relabelling
$t \leftrightarrow p$ and $J_1 J_2 \cdots J_s \leftrightarrow J_s \cdots  J_2 J_1$
applies to both the second and the third line of (\ref{ik.13}).

\subsubsection{Towards higher-genus Fay identities}

In order to find the primitive of 
\bea
\omega_I(1)\partial_{2} {\cal G}(2,1) \partial_{3} {\cal G}(3,1)
= \omega_I(z_1)\partial_{z_2} {\cal G}(z_2,z_1) \partial_{z_3} {\cal G}(z_3,z_1)
\label{Fayex.01}
\eea
with respect to $z_1$, the interchange lemmas (\ref{polylog.31}) and (\ref{ik.13}) 
need to be augmented by higher-genus  generalizations of the genus-one Fay identity,
\begin{align}
\Omega(z_1,\alpha_1|\tau)  \Omega(z_2,\alpha_2|\tau)
&= \Omega(z_1,\alpha_1{+}\alpha_2|\tau) \Omega(z_2{-}z_1,\alpha_2|\tau)\label{polylog.63} \\
&\quad
+\Omega(z_2,\alpha_1{+}\alpha_2|\tau) \Omega(z_1{-}z_2,\alpha_1|\tau)
\notag
\end{align}
of the Kronecker-Eisenstein series (\ref{polylog.04}). The need for identities beyond
the scope of interchange lemmas can be seen from the appearance of $z_1$ in
two factors $\partial_{2} {\cal G}(2,1)$ and $\partial_{3} {\cal G}(3,1)$ of (\ref{Fayex.01}) 
which persists after trading $\partial_{i} {\cal G}(i,1)\omega_I(1)$ for $\partial_{1} {\cal G}(1,i)\omega_I(i)$
and $\Phi$-tensors via (\ref{polylog.31}). As a first example of higher-genus Fay identities, 
the expression (\ref{Fayex.01}) can be rewritten as,
\begin{align}
\omega_I(1)\partial_{2} {\cal G}(2,1) \partial_{3} {\cal G}(3,1)
&= \Big\{ {-} \partial_{1} {\cal G}(1,3) \partial_{2} {\cal G}(2,3) \omega_I(3) 
- \omega_K(3) \partial_1 {\cal G}^K(1,2) \omega_I(2)
\label{polylog.65} \\
& \quad \quad \quad + \omega_K(3) \partial_{1} \Phi^{KL}{}_{\! I}(1) \omega_L(2)
+ \omega_K(3) \partial_1 \Phi^K{}_{\! L}(1) \partial_2 \Phi^L{}_{\! I}(2) 
\notag \\
& \quad \quad \quad 
+(z_2\leftrightarrow z_3) \Big\}+ \omega_K(1) \partial_{2} \partial_{3} V^K_I(2,3)
\notag
\end{align}
such that each term on the right-hand side features only one $z_1$-dependent factor
and can be readily integrated over $z_1$ via higher-genus polylogarithms (\ref{polylog.51}).
In the last line of (\ref{polylog.65}), we encounter the convolution,
\begin{align}
\partial_x \partial_y V^K_I(x,y)
&=  \int_{\Sigma} d^2 z \, \partial_x {\cal G}(x,z) 
\bar \om^K(z) \om_I(z) \partial_y {\cal G}(y,z)
\label{polylog.66} \\
&=
 \partial_y \Phi^K{}_L(y)  \partial_x \Phi^L{}_I(x)
+ \partial_y \Phi^{KL}{}_I(y) \omega_L(x) 
- \partial_y {\cal G}^K(y,x) \omega_I(x)
\notag
\end{align}
which has been reduced to the integration kernels  of (\ref{polylog.39}) in passing to the second line. The derivation of (\ref{polylog.65}) is based on the fact that the first three lines of the right-hand side have the same anti-holomorphic derivative with respect to $z_1,z_2,z_3$ as the left-hand side.  Moreover, both sides of (\ref{polylog.65}) vanish upon multiplication by $\bar \omega^P(1)\bar \omega^Q(2)\bar \omega^R(3)$ and integrating all of $z_1,z_2,z_3$ over the surface which excludes the addition of holomorphic terms.

\sm

At genus one, (\ref{polylog.65}) reduces to the Fay identity,
\beq
f^{(1)}(z_1{-}z_2|\tau) f^{(1)}(z_2{-}z_3|\tau) + f^{(2)}(z_1{-}z_3|\tau) + {\rm cycl}(z_1,z_2,z_3)=0
\eeq
at the $\alpha_1^0 \alpha_2^0$ order of (\ref{polylog.63}) through the identifications
(\ref{redGtens}) as well as
\bea
\partial_x \partial_y V^K_I(x,y) \, \big|_{h=1} = f^{(2)}(x{-}y|\tau)
\eea
Similar Fay identities should reduce higher-rank analogues 
$\omega_I(1)\partial_{2} {\cal G}^{I_1\ldots I_s}(2,1)  \partial_{3} {\cal G}^{J_1\ldots J_{s'}}(3,1)$
of (\ref{Fayex.01}) to convolutions of Arakelov Green functions 
whose primitives with respect to $z_1$ are determined by the differential equations of 
higher-genus polylogarithms. This is expected since products of convolutions of
Arakelov Green functions span a rich function space
which offers multiple distinct expressions with the same anti-holomorphic derivatives as
$\omega_I(1)\partial_{2} {\cal G}^{I_1\ldots I_s}(2,1) \partial_{3} {\cal G}^{J_1\ldots J_{s'}}(3,1)$.
By iteratively simplifying these alternative solutions to the differential equations with respect to $\bar z_i$ 
via interchange lemmas and lower-rank Fay identities, the $z_1$-dependence can be 
eventually arranged to occur through a single factor of either 
$\omega_I(1), \partial_1 \Phi^{I_1\cdots I_r}{}_J(1)$ or 
$\partial_1 {\cal G}^{I_1\cdots I_s}(1,i)$ in each term.

\sm

The detailed form of higher-rank Fay identities at arbitrary genus and
their proof can be found in \cite{DHoker:2024ozn}.

\subsection{Higher-genus associators} 
\label{sec:assoc}

The Drinfeld associator for genus zero evaluates monodromy properties of solutions to the Knizhnik-Zamolodchikov equation with a connection  given in (\ref{2.JKZ}) \cite{Drinfeld:1989, Drinfeld:1991}. The associator was generalized to genus one in \cite{Calaque:2009, Enriquez:2014, Hain:2013} using the meromorphic connection $\cJ_{\text{E}}(z|\tau)$, and the associated Knizhnik-Zamolodchikov-Bernard equation,
\bea
\p_z F(z|\tau) = \cJ_{\text{E}} (z|\tau) F(z|\tau) 
\hskip 0.8in 
\cJ_{\text{E}} (z|\tau) = - dz\, {\tet_1(z+{\rm ad}_b|\tau) {\rm ad}_b \over \tet_1(z|\tau) \tet_1({\rm ad}_b|\tau)} (a) 
\label{jeconn}
\eea
Since the connection satisfies $\cJ_{\text{E}} (z+1|\tau)=\cJ_{\text{E}} (z|\tau)$ and $\cJ_{\text{E}} (z+\tau|\tau) = e^{-2 \pi i \, {\rm ad}_b} \cJ_{\text{E}} (z|\tau)$ the  functions $F(z|\tau), F(z+1|\tau)$ and $e^{2 \pi i b} F(z+\tau|\tau)$ satisfy the same differential equation in $z$. The solutions are normalized in \cite{Enriquez:2014} by their behavior as $z \to 0$. As a result, their Wronskians are given as follows,\footnote{The associators $\Phi_\mA,\Phi_\mB$ and $\Phi_{\mA_I} ,\Phi_{\mB_I} $ in this section are not to be confused with the modular tensors $\Phi^I{}_J(x)$ and $\Phi^{I_1\cdots I_r}{}_J(x)$ defined by (\ref{polylog.28}) and (\ref{polylog.29}).}
\bea
\Phi_\mA (\tau) & = & F(z|\tau)^{-1} F(z+1|\tau) 
\no \\
\Phi_\mB (\tau) & = & F(z|\tau)^{-1} e^{2 \pi i b} F(z+\tau|\tau)
\eea
are independent of $z$, as the notation $\Phi_\mA(\tau)$ and $\Phi_\mB(\tau)$ indeed suggests, and are referred to as the elliptic associators introduced in \cite{Enriquez:2014}. Equivalently, the solution $F(z|\tau)$ may be expressed in terms of the path-ordered exponential of the connection $\cJ_{\text{E}}$ in (\ref{jeconn}),
\bea 
F(z|\tau) = \left ( \text{P} \exp \int _{z_0} ^z \cJ_{\text{E}} (t|\tau) \right ) F(z_0|\tau)
\label{poeJE}
\eea
where $t$ is the integration parameter and $z_0$ is a reference point chosen such that $F$ satisfies the normalization as $z \to 0$ introduced in \cite{Enriquez:2014}. In terms of this formulation, we obtain the following expressions for the elliptic associators,
\bea
\label{4.ass1}
\Phi_\mA(\tau) & = & F(z|\tau) ^{-1} 
\left ( \text{P} \exp \int ^{z+1} _{z} \cJ_{\text{E}} (t|\tau) \right ) F(z|\tau) 
\no \\
\Phi_\mB(\tau) & = & F(z|\tau) ^{-1} e^{ 2 \pi i  b} 
\left ( \text{P} \exp \int ^{z+\tau} _{z} \cJ_{\text{E}}  (t|\tau) \right ) F(z|\tau) 
\eea
Using the properties of the path-ordered exponential, it may be verified immediately that both expressions
are independent of $z$. A parallel construction may be given in terms of the non-meromorphic but doubly periodic Brown-Levin connection $\cJ_{\text{BL}}$ defined in (\ref{polylog.06a}), in which case the factor $e^{2 \pi i b}$ should be omitted.  It is this non-meromorphic version that we shall propose to generalize to higher genus.

\sm

Based on the polylogarithms that we have constructed in (\ref{polylog.51}), we are led to a natural proposal for {\it higher-genus associators}, which generalize the elliptic associators reviewed above. They may be defined by evaluating the path-ordered exponential of (\ref{polylog.51}) around the homology cycles of the Riemann surface.  Those integrals, in turn, are generated by the path-ordered exponentials of  (\ref{polylog.51}) around the basis of homology cycles $\mA_I$ and $\mB_I$, and defined  in close analogy with the construction given in (\ref{4.ass1}). We introduce a slight variant of the generating function for higher-genus polylogarithms defined in (\ref{polylog.51}), 
\bea
F(x;p|\Omega) = \left ( \text{P} \exp \int ^x _{x_0} \cJ(t;p|\Omega) \right ) F(x_0;p|\Omega ) = \bGam(x,x_0;p|\Omega ) F(x_0;p|\Omega)
\eea
Here, we have exhibited the dependence on the moduli of the higher-genus Riemann surface $\Sigma$ in terms of the period matrix $\Omega$ so that the presentation is as close as possible to the genus-one case (\ref{poeJE}), and retained the dependence on the point $p$ explicitly. Since the connection $\cJ(t;p|\Omega)$ is single-valued on $\Sigma$, the functions $F(x;p|\Omega), F(x+\mA_I;p|\Omega)$ and $F(x+\mB_I;p|\Omega)$ satisfy the same differential equation $d_x F(x;p|\Omega) = \cJ(x;p|\Omega) F(x;p|\Omega)$. As a result, the following combinations are independent of $x$,
\bea
\Phi_{\mA_I} (p|\Omega) & = & 
F(x;p|\Omega)^{-1} 
\left ( \text{P} \exp \int ^{x+\mA_I} _{x} \cJ(t;p|\Omega) \right ) F(x;p|\Omega) 
\no \\
\Phi_{\mB_I} (p|\Omega) & = & 
F(x;p|\Omega)^{-1} 
\left ( \text{P} \exp \int ^{x+\mB_I} _{x} \cJ(t;p|\Omega) \right ) F(x;p|\Omega) 
\eea
A significant difference between the proposal for higher-genus associators made here and the elliptic associators of Enriquez is the dependence on the point $p$ on $\Sigma$. Also, it remains to be investigated whether our proposal satisfies the general axioms enunciated for associators, or whether these axioms can be relaxed and generalized to the case of higher-genus polylogarithms. Moreover, it would be interesting to relate our proposal to the operad-theory approach to higher-genus associators in \cite{Gonzalez:2020}. We shall return to these open questions in future work, see section \ref{sec:conc}.

\newpage

\section{Flat connection in the multiple variable case}
\setcounter{equation}{0}
\label{sec:multi}

In this section, we further generalize the construction of the flat connection and polylogarithms
on higher-genus Riemann surfaces to an arbitrary number of marked points, thereby generalizing the 
multi-variable genus-one polylogarithms of Brown and Levin to higher genera. This generalized flat connection may be formulated entirely in terms of the modular tensors needed for the single-variable case, just as was the case for genus one.

\sm

The genus-one connection $\cJ_{\rm BL}$ in (\ref{polylog.06a}) may be generalized to depend on $n$ additional marked points $ z_1, \cdots, z_n$ on the torus, and is then given as follows by  \cite{BrownLevin}, 
\bea
\label{eq:BLMultivariable}
\mathcal{J}_{\mathrm{BL}}( z_1, \cdots, z_n ; z | \tau)
& = &
\frac{\pi}{\operatorname{Im} \tau} \, (d z-d \bar{z}) \, b+d z \, \operatorname{ad}_b \Omega(z, {\rm ad}_b | \tau) \, a  
\nonumber \\ &&
+ dz \sum_{i = 1}^n \big( \Omega(z - z_i, {\rm ad}_b | \tau) 
- \Omega(z, {\rm ad}_b | \tau) \big) \, c_i
\eea
We now have a Lie algebra ${\cal L}_n$ that is freely generated by the elements $a, b$
of section \ref{sec:blpoly} and additional elements $c_1,\cdots,c_n$ associated with
the marked points $z_1,\cdots,z_n$. The generalization to higher genus is given by the following theorem.

{\thm
\label{6.mv}
The connection $\cJ_{\rm mv}$ is defined on a compact Riemann surface $\Sigma$ of arbitrary genus $h$ and with $n+1$ marked points $p, z_1 , \cdots, z_n \in \Sigma$ as follows,  
\bea
\cJ _{\rm mv} (z_1,  \cdots, z_n ; x, p) &= &
\cJ (x,p)  + \cJ_n (z_1, \cdots, z_n ; x, p ) 
\label{thm03.01}
\eea
The connection $\cJ(x,p)$ for the single-variable case was constructed in Theorem \ref{3.thm1} and is repeated here for convenience, while the multi-variable addition $\cJ_n$ is given by, 
\bea
\cJ(x,p) &= & -\pi d \bar{x} \, \bar{\omega}^I(x) \,b_I+\pi d x\, \mathcal{H}^I(x ; B)\, b_I+d x \,\Psi_I(x, p ; B) \,a^I 
\no \\
\cJ_n (z_1,  \cdots, z_n ; x, p )  &= & 
- dx \sum_{i=1}^n \Big ( \cH(x, z_i ; B)- \cH (x, p ; B) \Big ) \,c_i
\label{thm03.02}
\eea
The connection $\cJ_{\rm mv}$ takes values in a Lie algebra $\cL_{\rm mv}$ that is freely generated by the elements $a^I, b_I$ with $I=1,\cdots, h$ as seen in section \ref{sec:flcon} and additional elements $c_i$ with $i=1,\cdots , n$ associated with the marked points $z_i$. The connection $\cJ_{\rm mv}$ is flat away from the points $p, z_1, \cdots, z_n$,
\bea
d \cJ_{\rm mv} -\cJ_{\rm mv} \wedge \cJ_{\rm mv}
= \pi d \bar{x} \wedge d x
\bigg (\delta(x, p) \Big (\left[b_I, a^I\right]-\sum_{i=1}^n c_i \Big )+\sum_{i=1}^n \delta (x, z_i )  \, c_i \bigg )
\label{thm03.03}
\eea
and reduces to the multi-variable Brown-Levin connection (\ref{eq:BLMultivariable}) in the genus-one case. Alternatively, one may express the connection $\cJ(x,p)$ in terms of the basis $(\hat a^I, b_I)$
with tensorial modular transformations. This leads to a modular invariant multi-variable
connection (\ref{thm03.01}) if the generators $c_i$ are taken to be $Sp(2h,\mathbb Z)$-invariant.} 

\sm

To prove the theorem, we begin by showing that the connection $\cJ_{\rm mv}$ in (\ref{thm03.01}) and (\ref{thm03.02}) reduces to $\cJ_\text{BL}$ for $h=1$. Indeed, the differences between the generating series $\cH$ in $\cJ_n$ reduce to differences between the  Kronecker-Eisenstein series  inside the sum in (\ref{eq:BLMultivariable}). In particular, by expanding the components, it is easy to see from (\ref{redGtens}) that we have,
\bea
\cH(x, z_i ; B)-\cH(x, p ; B) \, \big |_{h=1} = 
-\Omega (x-z_i, {\rm ad}_b | \tau)
+\Omega (x, {\rm ad}_b | \tau )
\eea
upon setting $p \to 0 $ on the torus and identifying $a^1=a$ as well as $b_1=b$.

\sm

Furthermore, to prove flatness of $\cJ_{\rm mv}$ away from the points $y,z_i$, we use the flatness condition
of the connection $\cJ$ in Theorem \ref{3.thm1} to obtain the following relation, 
\bea
d \cJ_{\rm mv} - \cJ_{\rm mv} \wedge \cJ_{\rm mv} = d \cJ_n - \cJ \wedge \cJ_n - \cJ_n \wedge \cJ
+ \pi d \bar x \wedge dx \, \delta(x,p) [b_I, a^I]
\eea
The remaining contributions may be worked out using (\ref{polylog.36}) as follows, 
\bea
d \cJ_n & = & -d\bar{x} \wedge d x \sum_{i=1}^n
\Big ( \pbx \cH(x, z_i ; B)- \pbx \cH(x, p ; B) \Big ) \, c_i
\no \\
& = &
\pi d\bar{x} \wedge d x \sum_{i=1}^n \Big (  \delta(x, z_i) - \delta(x, p)+ \bar{\omega}^I(x) B_I 
\big ( \cH(x, z_i ; B) - \cH(x, p ; B) \big ) \Big ) \, c_i
\quad
\eea
Similarly,  the wedge products may be worked out as follows,
\bea
\cJ \wedge \cJ_n + \cJ_n \wedge \cJ = \pi d\bar{x} \wedge dx \, \bar{\omega}^I(x) 
\bigg [ b_I , \sum_{i=1}^n \Big (\cH(x, z_i ; B)-\cH(x, p ; B) \Big ) \, c_i \bigg ]
\eea
Flatness of $\cJ_{\rm mv}$ then follows in view of $B_I  \cH(x, p ; B) c_i = [b_I,  \cH(x, p ; B) c_i]$.

\sm

Finally, modular invariance of $\cJ (x,p), \cH(x, z_i ; B){-}\cH(x, p ; B)$ and $c_i$ readily carries over
to the multi-variable connection $ \cJ_{\rm mv}$ in (\ref{thm03.01}) and (\ref{thm03.02}).

\sm

Multi-variable generalizations of the higher-genus polylogarithms in section \ref{sec:hgenpoly} may now be constructed from the flat connection $\cJ_{\rm mv}$ following the methods presented in section \ref{sec:polylogs}.

\newpage

\section{Separating degeneration}
\label{sec:sepdeg}
\setcounter{equation}{0}

In this section, we shall study the behavior of the generating functions and flat connection
at higher genus under separating degenerations of the genus-$h$ Riemann surface~$\Sigma$. To keep the discussion as simple and concrete as possible, we shall specialize here to the case of genus two degenerating to two genus-one surfaces connected by a long funnel. The generalization of this construction to the non-separating  degeneration of higher-genus Riemann surfaces may be found in section 3 of \cite{DHoker:2017pvk}, while the separating, non-separating, and tropical degenerations for  genus two are presented in detail in \cite{DHoker:2018mys}.

\subsection{The construction for genus two}

\begin{figure}[h]
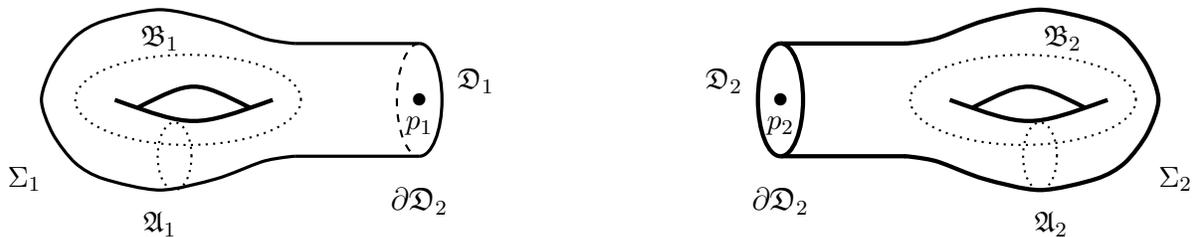

\begin{center}
\tikzpicture[scale=1.5]
\scope[xshift=-4cm,yshift=0cm]
\draw [ultra thick] (0.5,-1) arc (-90:90:0.2 and 0.5);
\draw [ultra thick] (0.5,0) arc (90:270:0.2 and 0.5);
\draw [ultra thick] (0.5,0) -- (1.61,0);
\draw [ultra thick] (0.5,-1) -- (1.61,-1);
\draw[ultra thick] plot [smooth] coordinates {
(1.6,0) (1.7, 0.01) (1.9, 0.05) (2.3, 0.2) (2.8, 0.3) (3.3, 0.2) (3.55, 0.06) (3.75, -0.2)   
(3.85,-0.5) 
(3.75,-0.8) (3.55, -1.06) (3.3,-1.2) (2.8, -1.3) (2.3, -1.2) (1.9, -1.05) (1.7, -1.01)(1.6,-1)};
\draw[ultra thick] (2,-0.5) .. controls (2.7, -0.75) .. (3.4,-0.5);
\draw[ultra thick] (2.2,-0.57) .. controls (2.7, -0.32) .. (3.2,-0.57);
\draw [thick, dotted] (3.65,-0.5) arc (0:360:1 and 0.4);
\draw [thick, dotted] (2.95,-1) arc (0:360:0.15 and 0.3);
\draw [very thick] (-2.7,-1) arc (-90:90:0.2 and 0.5);
\draw [thick, dashed] (-2.7,0) arc (90:270:0.2 and 0.5);
\draw [very thick] (-2.7,0) -- (-3.8,0);
\draw [very thick] (-2.7,-1) -- (-3.8,-1);
\draw[very thick] plot [smooth] coordinates {
(-3.8,0) (-3.9, 0.01) (-4.1, 0.05)(-4.5, 0.2) (-5, 0.3) (-5.5, 0.2) (-5.75, 0.06) (-5.95, -0.2)   
(-6.05,-0.5) 
(-5.95,-0.8) (-5.75, -1.06) (-5.5,-1.2) (-5, -1.3) (-4.5, -1.2) (-4.1, -1.05) (-3.9, -1.01)(-3.8,-1)};
\draw[ultra thick] (-4,-0.5) .. controls (-4.7, -0.75) .. (-5.4,-0.5);
\draw[ultra thick] (-4.2,-0.57) .. controls (-4.7, -0.32) .. (-5.2,-0.57);
\draw [thick, dotted] (-3.75,-0.5) arc (0:360:1 and 0.4);
\draw [thick, dotted] (-4.72,-1) arc (0:360:0.15 and 0.3);
\draw (0.5,-1.4) node{$\p \mD_2$};
\draw (0.5,-0.5) node{$\bullet$};
\draw (0.5,-0.73) node{{\small $p_2$}};
\draw (0,-0.33) node{{\small $\mD_2$}};
\draw (3,0.05) node{{\small $\mB_2$}};
\draw (2.9,-1.6) node{{\small $\mA_2$}};
\draw (4,-1.2) node{{\small $\Sigma_2$}};
\draw (-2.7,-1.4) node{$\p \mD_1$};
\draw (-2.7,-0.5) node{$\bullet$};
\draw (-2.7,-0.73) node{{\small $p_1$}};
\draw (-2.2,-0.33) node{{\small $\mD_1$}};
\draw (-5,0.05) node{{\small $\mB_1$}};
\draw (-5,-1.6) node{{\small $\mA_1$}};
\draw (-6.2,-1.2) node{{\small $\Sigma_1$}};
\endscope
\endtikzpicture
\caption{Funnel construction of a family of genus-two Riemann surfaces~$\Sigma$ near the separating divisor  in terms of genus-one surfaces $\Sigma _1$ and $\Sigma _2$. The circles  $\p \mD_1$ and $\p \mD_2$  are centered at the punctures $p_1$ and $p_2$ and bound the  discs $\mD_1$ and $\mD_2$, respectively. The surface $\Sigma$ is constructed from the surfaces $\Sigma _1 \setminus \mD_1$  and $\Sigma _2 \setminus \mD_2$ by identifying $\p \mD_1$ and $\p \mD_2$.}
\label{fig:funnel}
\end{center}
\end{figure}

A convenient parametrization of the neighborhood of the separating divisor is provided by the funnel construction given in Fay's book \cite{Fay:1973}, and specifically for genus two in \cite{DHoker:2018mys}. Here we shall give a simplified presentation that will suffice for the problem at hand. For genus two, the starting point of the construction of $\Sigma$  is provided by the compact genus-one surfaces $\Sigma _1$ and $\Sigma_2$, to which we add punctures $p_1$ and~$p_2$, respectively. Next, we introduce a system of local complex coordinates $(x_1,\bar x_1)$ and $(x_2, \bar x_2)$ on each surface, and denote the coordinates of the punctures simply by $p_1$ and~$p_2$. We specify a disc $\mD_1$  centered at $p_1$ on $\Sigma_1$ and a disc $\mD_2$ centered at $p_2$ on $\Sigma_2$, as shown in Figure \ref{fig:funnel}.

The genus-two surface $\Sigma$ is obtained by identifying annuli surrounding $\p \mD_1$ and $\p \mD_2$ with respective local complex coordinates  $x_1$ and $x_2$ via the relation (more complete discussions of the construction  are given in \cite{Fay:1973,DHoker:2018mys}),
\bea
\label{fay1}
(x_1-p_1)(x_2-p_2) = v_s
\eea
Here $v_s$ is a complex parameter governing the separating degeneration (which is referred to as $t$ in \cite{Fay:1973}) and is such that the separating degeneration corresponds to the limit $v_s \to 0$. Customarily, the curves $\p \mD_1, \p \mD_2$ are defined to be circles in the local complex coordinates on the surfaces but here instead we shall use a more intrinsic definition, 
\begin{align}
\p \mD_1 & =  \{ x_1 \in \Sigma _1 \hbox{ such that } g(x_1-p_1|\tau)=t_1 \}
\no \\
\p \mD_2 & =  \{ x_2 \in \Sigma _2 \hbox{ such that } g(x_2-p_2|\sigma)=t_2 \}
\label{modd1d2}
\end{align}
where the scalar Green function $g(z|\tau)= g_1(z|\tau)$ on the torus was defined in (\ref{2.green}).
The moduli $\tau$ and $\sigma$ of the genus-one surfaces $\Sigma_1$ and $\Sigma_2$
in (\ref{modd1d2}) are given by the diagonal entries of the period matrix $\Omega$ of 
the genus-two Riemann surface $\Sigma$ which will be parametrized by,
\bea
\Omega = \left ( \begin{matrix} \tau & v \cr v & \sigma \cr \end{matrix} \right )
\label{permat}
\eea
Note that, for sufficiently large values of $t_1, t_2$, each level-set $\p \mD_1, \p \mD_2$ in (\ref{modd1d2}) is connected.

\sm

In the separating degeneration limit, the diagonal entries $\tau, \sigma$ are kept fixed as $v \to 0$. The relation between the off-diagonal entry $v$ and $v_s$ is linear and will be derived shortly in (\ref{vtovs}) below. The relation between the parameters $t_1, t_2$, and $v_s$ is obtained as follows,
\bea
t_1+t_2= - \ln |2 \pi  v_s \, \eta (\tau)^2 \eta (\sigma )^2|^2 + \cO(v_s^2)
\eea
Here, we have used the short-distance expansion of the scalar Green function on the torus, 
\bea
 g(z|\tau) = - \ln | 2 \pi z \, \eta (\tau)^2|^2 + \cO(z^2)
 \label{asympgf}
 \eea
 to convert  (\ref{fay1}) into the  expression above.

\sm

When performing integrals over the genus-two surface $\Sigma$, it will be convenient to 
decompose the integral into a sum of the contribution from $\Sigma_1 \setminus \mD_1$ plus the contribution from $\Sigma_2 \setminus \mD_2$, where the curves $\p \mD_1, \p \mD_2$ are defined such that $t_1=t_2=t$. Under these conditions, the Abelian differentials $\om_1$ and $\om_2$ remain uniformly bounded throughout $\Sigma$ by a constant of order $\cO(v_s^0)$, with corrections which are suppressed by powers of $v_s$.

\sm

Assuming that we set $t_1=t_2$, we can evaluate the dependence on $v$ of the radii of the coordinate discs $\mD_1, \mD_2$ in the limit where $v$ is small. Using the asymptotic expression (\ref{asympgf}) for the Green functions, we have for $x_1 \in \mD_1$ and $x_2 \in \mD_2$, 
\bea
\label{6.dist}
 |x_1-p_1| = |x_2-p_2|= |2 \pi v_s \eta (\tau)^2 \eta(\sigma)^2|^\half
\eea
Thus, in the limit where $v_s \to 0$, the coordinate areas of the coordinate discs tend to zero linearly in $v_s$ and thus linearly in $v$, as we shall establish below.

\subsection{Degeneration of Abelian differentials}

We choose  canonical homology bases  $\mA_1, \mB_1 \subset \Sigma _1 \setminus \{p_1\}$ and $\mA_2,\mB_2 \subset \Sigma _2 \setminus \{p_2\}$ as in Figure~\ref{fig:funnel}, and extend those to a canonical homology basis for $\Sigma$. The genus-one holomorphic Abelian differentials  $\hat \omega_1$ and $\hat  \omega_2$ on  $\Sigma_1$ and $\Sigma_2$, respectively, are normalized  as follows, 
\bea
\oint _{\mA_1} \hat \om_1 = \oint _{\mA_2} \hat \om_2 = 1
\hskip 1in 
\oint _{\mB_1} \hat \om_1 = \tau 
\hskip 1in 
\oint _{\mB_2} \hat \om_2 = \sigma 
\eea
To construct holomorphic one-forms on the genus-two surface $\Sigma$ with the period matrix $\Omega$ parametrized in (\ref{permat}) we extend $\hat \om_1$ to a differential $\om_1$ on $\Sigma$ and $\hat \om_2$ to a differential $\omega_2$ on $\Sigma$ by using the identification (\ref{fay1}). Choosing  complex coordinates $x_1, x_2$ on $\Sigma _1$ and $\Sigma _2$  such that $\hat \om_1= dx_1$ and $\hat \om_2 = dx_2$, we see that the differential $dx_1$ extends to $- v_s /(x_2-p_2)^2dx_2 $ in $\Sigma _2$ while the differential $dx_2$ extends to $- v_s /(x_1-p_1)^2dx_1$ in $\Sigma _1$. Thus, the extensions are governed by meromorphic one-forms with a double pole. The meromorphic one-forms  $\varpi (x_1,y_1|\tau)$ and $\varpi(x_2,y_2| \sigma)$ on $\Sigma _1$ and $\Sigma _2$, respectively, are normalized to have vanishing $\mA$-periods and a double pole of unit strength  at $x_1=y_1$ and $x_2=y_2$. Their $\mB$-periods are given by the Riemann bilinear relations,
\bea
\oint _{\mB_1} \varpi(x_1,y_1|\tau) = 2 \pi i \hat \om_1(y_1)
\hskip 1in 
\oint _{\mB_2} \varpi (x_2,y_2|\sigma) = 2 \pi i \hat \om_2 (y_2)
\eea
The holomorphic one-forms $\om_1$ and $\om_2$ on the genus-two surface $\Sigma$, canonically normalized on $\mA_1$ and $\mA_2$-cycles, are then given as follows,\footnote{Here and below, we shall use the notation $x$ for a point on the genus-two surface $\Sigma$, and set it equal to $x_1$ when the point lies in the genus-one component $\Sigma _1 \setminus \mD_1$ and to $x_2$ when it lies in the component $\Sigma _2 \setminus \mD_2$.}
\bea
\label{SDAb}
\omega_1(x) & \to & \begin{cases}
\hat \omega_1 (x_1)& ~~~ x_1 \in \Sigma_1 \setminus \mD_1 
\\
v  \,  \varpi (x_2,p_2|\sigma ) / ( 2 \pi i \hat \om_2 (p_2))  ~~  & ~~~ x_2\in\Sigma_2 \setminus \mD_2
\end{cases}
\no \\
\omega_2(x) & \to & \begin{cases}
 v \,  \varpi(x_1,p_1|\tau) /  (2 \pi i \hat \omega_1(p_1) ) ~~ & ~~~  \, x_1\in\Sigma_1 \setminus \mD_1  
 \\
\hat \omega_2(x_2) ~~~ & ~~~ \, x_2\in \Sigma_2 \setminus \mD_2
\end{cases}
\eea
The parameter $v_s $ is related to the off-diagonal entry $v$ of the genus-two period matrix $\Omega$ of (\ref{permat}) by, 
\bea
v = \oint _{\mB_1} \om _2 = \oint _{\mB_2} \om _1 = - 2 \pi i \, v_s \, \hat \om_1(p_1) \, \hat \om _2(p_2)
\label{vtovs}
\eea
The expressions  in (\ref{SDAb}) are valid up to corrections of order $\cO(v^2)$ which have been omitted.

\subsection{Degeneration of the Arakelov Green function}

The separating degeneration of the Arakelov Green function $\cG$ on the funnel construction
of genus-two surfaces $\Sigma$ in Figure \ref{fig:funnel} is given by, 
\bea
\label{SDArak}
\cG(x,y) \to \begin{cases}
- \half \ln |\hat v|  + g(x_1-y_1|\tau) - \half g(x_1-p_1|\tau) - \half g(y_1-p_1|\tau) 
   \qquad x_1, y_1 \in \Sigma _1 \setminus \mD_1 
\no   \\
- \half \ln |\hat v|  + g(x_2-y_2|\sigma) - \half g(x_2-p_2|\sigma ) - \half g(y_2-p_2|\sigma ) 
  \qquad x_2, y_2 \in \Sigma _2 \setminus \mD_2
  \\
+ \half \ln |\hat v| +\half g(x_1-p_1|\tau)+\half g(y_2-p_2|\sigma) 
   \qquad \quad x_1 \in\Sigma_1 \setminus \mD_1, \,   y_2 \in\Sigma_2 \setminus \mD_2
    \\
+ \half \ln |\hat v| +\half g(x_2-p_2|\sigma)+\half g(y_1-p_1|\tau) 
   \qquad \quad x_2 \in\Sigma_2 \setminus \mD_2, \,   y_1 \in\Sigma_1 \setminus \mD_1
\end{cases}
\no \\
 \eea
where $\hat v$ is related to the entries $v,\tau,\sigma$ of the
genus-two period matrix (\ref{permat}) by the Dedekind eta-function $\eta$,
\bea
\hat v = 2 \pi \, v \, \eta (\tau)^2 \eta (\sigma )^2
\eea
The combinations that will enter the flat connection and the generating functions
of genus-two polylogarithms are,
\bea
\label{Arakder}
\p_x \cG(x,y) \to \begin{cases}
-f^{(1)} (x_1-y_1|\tau) + \half f^{(1)} (x_1-p_1|\tau) 
   & x_1, y_1 \in \Sigma _1 \setminus \mD_1 
   \\
-f^{(1)} (x_2-y_2|\sigma) + \half f^{(1)} (x_2-p_2|\sigma )
  & x_2, y_2 \in \Sigma _2 \setminus \mD_2
  \\
- \half f^{(1)}  (x_1-p_1|\tau)
   & x_1 \in\Sigma_1 \setminus \mD_1 \, , \, y_2 \in\Sigma_2 \setminus \mD_2
   \\
 - \half f^{(1)}  (x_2-p_2|\sigma)
   &x_2 \in\Sigma_2 \setminus \mD_2\, , \,   y_1 \in\Sigma_1 \setminus \mD_1
\end{cases}
\qquad \eea
Conveniently, the dependence on $\ln |\hat v|$ cancels out in these derivative combinations.

\subsection{Degeneration of $\p_x \cG^{I_1 \cdots I_s}(x,y)$}

We begin with the separating degenerations of the convolutions $\p_x \cG^{I_1 \cdots I_s}(x,y)$ of Arakelov Green functions defined in (\ref{polylog.29}) that carry the dependence of the flat connection
on the second marked point $y$, 
\bea
\label{6.decomp}
\p_x \cG^{I_1 \cdots I_s} (x,y) & = & 
\int  _\Sigma d^2 z \, \p_x \cG(x,z) \oom^{I_1} (z) \p_z \cG^{I_2 \cdots I_s} (z,y) 
\no \\ 
& = & \p_x \cG^{I_1 \cdots I_s} _{(1)} (x,y) + \p_x \cG^{I_1 \cdots I_s} _{(2)}(x,y) 
\eea
where the second equality is obtained by decomposing  $\Sigma = \big ( \Sigma _1 \setminus \mD_1 \big ) \cup \big ( \Sigma _2 \setminus \mD_2 \big )$ with,
\bea
\label{G12}
\cG^{I_1 \cdots I_s} _{(i)}(x,y) = \int  _{\Sigma_i \setminus \mD_i}  \!\!\!\! d^2 z_i \, \cG(x,z_i) \, \oom^{I_1} (z_i) 
\, \p_{z_i}  \cG^{I_2 \cdots I_s} (z_i,y)
\hskip 0.8in i=1,2
\eea
The asymptotics is given by the following lemma.

{\lem 
\label{6.thm3}
The functions $\p_x \cG^{I_1 \cdots I_s} (x,y)$ have the following separating degenerations, up to order $\cO(v)$ which vanishes as $v \to 0$, 
\begin{align}
 \p_{x_1} \cG^{1_s} (x_1, y_1 ) &\to -f^{(s+1)} (x_1-y_1|\tau) + \thalf f^{(s+1)} (x_1-p_1|\tau)
\no \\
 \p_{x_1} \cG^{1_s} (x_1, y_2 ) &\to  - \thalf f^{(s+1)} (x_1-p_1|\tau)
\no \\
 \p_{x_1} \cG^S(x_1, y_i ) &\to  0 \hskip 0.8in {\rm if} \ 2 \in S \hskip 0.8in i=1,2
\no \\ 
\no \\
 \p_{x_2} \cG^{2_s} (x_2, y_2 ) &\to -f^{(s+1)} (x_2-y_2|\sigma) 
+ \thalf f^{(s+1)}  (x_2-p_2|\sigma)
\no \\
 \p_{x_2} \cG^{2_s} (x_2, y_1 ) &\to - \thalf f^{(s+1)} (x_2-p_2|\sigma)
\no \\
 \p_{x_2} \cG^S(x_2, y_i ) &\to  0
  \hskip 0.8in {\rm if} \ 1 \in S \hskip 0.8in i=1,2
 \label{6.thm3a}
\end{align}
where $1_s=1\cdots 1$ with $s$ entries and similarly for $2_s$, and where $S$ stands for an arbitrary  array $I_1 \cdots I_s$ with each $I_i$ taking values 1 or 2.}

\sm

The proof proceeds by induction on $s$. For $s=0$, only the cases on the first, second, fourth and fifth lines arise, and they are given by (\ref{Arakder}). We shall prove the validity of the formulas of (\ref{6.thm3a}) for $s=1$ in Appendix \ref{sec:sepapp}, as these cases involve some detailed analyses. The recursive definitions of (\ref{G12}) may then be used to show that the $v \to 0$ limit of any tensor $\p_x \cG^{I_1 \cdots I_s}(x,y)$ vanishes unless $I_1= \cdots = I_s=1$ or $I_1=\cdots I_s=2$. In these cases, it is readily shown that $\p_x \cG_{(2)}^{1_s}(x,y)$ and $ \p_x \cG_{(1)}^{2_s}(x,y)$ have vanishing limits, while $\p_x \cG_{(1)}^{1_s}(x,y)$ and $ \p_x \cG_{(2)}^{2_s}(x,y)$ produce the remaining contributions to (\ref{6.thm3a}).

\subsection{Degeneration of $\p_x \Phi^{I_1 \cdots I_s }{}_J(x)$}

It will be convenient to rewrite the recursive definition (\ref{polylog.29}) of the functions $\p_x \Phi^{I_1 \cdots I_s }{}_J(x)$ in terms of 
$\p_x \cG^{I_1 \cdots I_{s-1}}(x,y)$ whose separating degeneration was already evaluated in the preceding subsection, 
\bea
\p_x \Phi^{I_1 \cdots I_s }{}_J(x) = \int _\Sigma d^2 z \, \p_x \cG^{I_1 \cdots I_{s-1}} (x,z) \, \bar \om^{I_s}(z) \om_J(z)
\eea 
One may split up the integration into the contributions from the surfaces $\Sigma_1\setminus \mD_1$ and $\Sigma_2\setminus \mD_2$,  
\bea
\p_x \Phi^{I_1 \cdots I_s }{}_J(x) = 
 \big ( \p_x \Phi_{(1)} \big ) ^{I_1 \cdots I_s }{}_J(x)  +  \big ( \p_x \Phi_{(2)} \big ) ^{I_1 \cdots I_s }{}_J(x)
\eea
where,
\bea
\big ( \Phi_{(i)} \big ) ^{I_1 \cdots I_s} {}_J (x) 
= \int  _{\Sigma_i\setminus \mD_i}  d^2 z_i \, \cG^{I_1 \cdots I_{s-1}} (x,z_i) \, 
\bar \om ^{I_s} (z_i) \, \om _J(z_i) \hskip 1in i=1,2
\eea
The analysis parallels the one carried out for $\p_x \cG^{I_1 \cdots I_s}(x,y)$, uses the results of Lemma \ref{6.thm3}, and produces the following lemma, stated without proof.

{\lem 
\label{6.thm4}
The separating degenerations of the functions $\p_x \Phi^{I_1 \cdots I_s }{}_J(x)$ are given by,
\begin{align}
 \p_{x_1} \Phi ^{1_s} {}_1 (x_1) & \to  \thalf f^{(s)} (x_1-p_1|\tau)  
 & \qquad
 \p_{x_1} \Phi^{1_{s-1} 2}{}_2 (x_1) & \to  - \thalf f^{(s)} (x_1-p_1|\tau) 
 \no \\
 \p_{x_2} \Phi ^{2_s} {}_2 (x_2) & \to  \thalf f^{(s)} (x_2-p_2|\sigma) 
 &
 \p_{x_2} \Phi^{2_{s-1} 1}{}_1 (x_2) &\to  - \thalf f^{(s)} (x_2-p_2|\sigma) 
 \label{phisepdeg}
\end{align}
where again $1_s=1\cdots 1$ with $s$ entries and similarly for $2_s$. All 
other components such as  $\p_{x_i} \Phi ^{\cdots 1} {}_2 (x_i),
\p_{x_i} \Phi ^{\cdots 2} {}_1 (x_i)$ with $i=1,2$ or $ \p_{x_2} \Phi ^{1_s} {}_1 (x_2)$ with
$s\geq 2$ vanish. The degenerations of (\ref{phisepdeg}) are consistent
with the traceless property $\Phi^{I_1\cdots I_s}{}_{I_s}=0$.}

\subsection{Degeneration of the flat connection}
\label{sec:sepdegJ}

The separating degenerations of the modular tensors $\p_x \cG^{I_1 \cdots I_s}(x,y)$ and $\p_x \Phi^{I_1 \cdots I_s}{}_J(x)$ for the special case of genus-two surfaces  may now be used to evaluate the separating degeneration limit of the flat connections discussed in Theorem \ref{3.thm1} and Theorem \ref{6.mv}. 

\sm

We begin by assembling the limits of the generating function $\cH(x,y;B)$ in the first line of (\ref{polylog.35}) with $B_I = {\rm ad}_{b_I}$, 
\bea
\cH(x_1,y_1) & \to & \sum _{n=1}^\infty \Big ( {-} f^{(n)} (x_1-y_1|\tau) + \thalf f^{(n)} (x_1-p_1|\tau) \Big ) B_1^{n-1}
\no \\
\cH(x_2,y_2) & \to & \sum _{n=1}^\infty \Big ( {-} f^{(n)} (x_2-y_2|\sigma) + \thalf f^{(n)} (x_2-p_2|\sigma) \Big ) B_2^{n-1}
\no \\
\cH(x_1,y_2) & \to & - \half \sum _{n=1}^\infty  f^{(n)} (x_1-p_1|\tau) B_1^{n-1}
\no \\
\cH(x_2,y_1) & \to & - \half \sum _{n=1}^\infty  f^{(n)} (x_2-p_2|\sigma) B_2^{n-1}
\eea
as well as the components $I=1,2$ of $\cH_I(x;B)$ in the second line of (\ref{polylog.35}),
\bea
\cH_1(x_1) & \to & 1 + \half \sum_{n=1}^\infty f^{(n)} (x_1-p_1|\tau) B_1^n
\no \\
\cH_2(x_2) & \to & 1 + \half \sum_{n=1}^\infty f^{(n)} (x_2-p_2|\sigma) B_2^n
\no \\
\cH_1(x_2) & \to & - \half \sum_{n=1}^\infty f^{(n)} (x_2-p_2|\sigma) B_2^{n-1} B_1
\no \\
\cH_2(x_1) & \to & - \half \sum_{n=1}^\infty f^{(n)} (x_1-p_1|\tau) B_1^{n-1} B_2
\label{deghvec}
\eea
Note that the rightmost generator in both of $\cH_I(x_1)$ and $\cH_I(x_2)$ is given by $B_I$ in each term of (\ref{deghvec}). As a consequence, the contribution  $\sim \cH^I (x;B) \, b_I$ to the flat connection (\ref{polylog.39}) at genus two  reduces to $\omega^I(x) b_I$ under separating degenerations in view of $Y^{IJ} B_I b_J=0$.

\sm

By the composition (\ref{polylog.37}) of $\Psi_I(x,y)$, these results immediately imply the vanishing of two cases
for its components and arguments,
\bea
\Psi_1(x_2,y_1) & \to &0
\no \\
\Psi_2(x_1,y_2) & \to &0
\label{psidegs1}
\eea
Assembling the components $\Psi_I(x,y)$ with non-vanishing limits  we find, 
\bea
\Psi_1 (x_1,y_1) & \to & 1+ \sum _{n=1}^\infty f^{(n)} (x_1-y_1|\tau) B_1^n 
= B_1 \, \Omega(x_1-y_1, B_1|\tau)
\no \\
\Psi_2 (x_2,y_2) & \to & 1+ \sum _{n=1}^\infty f^{(n)} (x_2-y_2|\sigma) B_2^n 
= B_2 \, \Omega(x_2-y_2, B_2|\sigma)
\no \\
\Psi_1(x_1,y_2) & \to & 1+ \sum _{n=1}^\infty f^{(n)} (x_1-p_1|\tau) B_1^n 
= B_1 \, \Omega(x_1-p_1, B_1|\tau)
\no \\
\Psi_2(x_2,y_1) & \to & 1+ \sum _{n=1}^\infty f^{(n)} (x_2-p_2|\sigma) B_2^n 
= B_2 \,  \Omega(x_2-p_2, B_2|\sigma)
\no \\
\Psi_1(x_2,y_2) & \to & \sum _{n=1}^\infty \Big ( f^{(n)} (x_2-y_2|\sigma) - f^{(n)} (x_2-p_2|\sigma) \Big ) B_2^{n-1} B_1 
\no \\
&& \qquad = \big( \Omega(x_2-y_2,B_2|\sigma) - \Omega(x_2-p_2,B_2|\sigma) \big) B_1
\no \\
\Psi_2(x_1,y_1) & \to & \sum _{n=1}^\infty \Big ( f^{(n)} (x_1-y_1|\tau) - f^{(n)} (x_1-p_1|\tau) \Big ) B_1^{n-1} B_2 \no \\
&& \qquad = \big( \Omega(x_1-y_1,B_1| \tau ) - \Omega(x_1-p_1,B_1|\tau) \big) B_2
\label{psidegs}
\eea
Our results may be summarized in the form of the following theorem.

\sm

{\thm
To evaluate the separating degeneration of the flat connection in Theorem~\ref{3.thm1} at genus two, we may pick, without loss of generality, a point $y = y_1 \in \Sigma _1 \setminus \mD_1$. The components of the connection then enjoy the following asymptotics, 
\begin{align}
\cJ(x_1, y_1 ) & \to  
  { \pi \over \Im \tau} \,(dx_1 - d \bar x_1) \, b_1 +
dx_1 \, {\rm ad}_{b_1} \Omega \big (x_1-y_1, {\rm ad}_{b_1}|\tau \big ) \,  a^1 
\no \\ 
&\qquad
+ dx_1 \, \Big( \Omega(x_1-y_1, {\rm ad}_{b_1}|\tau) - \Omega(x_1-p_1, {\rm ad}_{b_1}|\tau) \Big)
\, [b_2,a^2] 
\no \\
\cJ(x_2, y_1 ) & \to
 { \pi \over \Im \sigma} \, (dx_2 - d \bar x_2) \, b_2 + dx_2 \, {\rm ad}_{b_2} 
 \Omega \big (x_2-p_2, {\rm ad}_{b_2} |\sigma \big ) \, a^2 
\label{thmsep}
\end{align}}

\sm

The proof proceeds by inserting (\ref{deghvec}), (\ref{psidegs1}) and (\ref{psidegs}) into the genus-two instance of the flat connection (\ref{polylog.39}).

\sm

We conclude this section with a number of remarks.
\begin{enumerate}
\itemsep=0in
\item Upon choosing $y_1=p_1$ the second line in the expression (\ref{thmsep}) for $\cJ(x_1,y_1)$ cancels. With this assumption, the connection $\cJ(x,y)$ with $y=y_1=p_1=0$ reduces to the genus-one Brown-Levin connection (\ref{polylog.06b}) on a torus $\Sigma_1$ when the point $x=x_1$ is on the surface  $\Sigma_1 \setminus \{p_1\}$. 
\item At distinct points $y_1\neq p_1$ and $y_1=0$, in turn, the separating degeneration of 
$\cJ(x_1, y_1 )$ in (\ref{thmsep}) yields the Brown-Levin connection (\ref{eq:BLMultivariable}) 
with one extra marked point $z_1=p_1$, and free Lie algebra generator $c_1= [a^2,b_2]$.
\item Finally, when the point $x=x_2$ is on the surface $\Sigma _2 \setminus \{ p_2\}$,
the genus-two connection $\cJ(x_2, y_1 )$ reduces to the Brown-Levin connection
on $\Sigma_2$ with a shift by $p_2$ in the Kronecker-Eisenstein series of (\ref{polylog.06b}).
\item The same methods determine the separating degeneration of the multi-variable addition 
${\cal J}_n(z_1,\cdots,z_n;x,y)$ to the flat connection at genus two in (\ref{thm03.02}). Once
the extra punctures are distributed via $z_1,z_2,\cdots ,z_m \in \Sigma _1 \setminus \mD_1$
and $z_{m+1},\cdots ,z_n \in \Sigma _2 \setminus \mD_2$, there are two inequivalent cases to
consider: for $x=x_1$ and $y=y_1$ on the same surface $ \Sigma _1 \setminus \mD_1$,
we obtain the multi-variable additions (\ref{eq:BLMultivariable}) to the genus-one connection
\begin{align}
\cJ_n(z_1,\cdots,z_n;x_1, y_1 ) &\to dx_1 \sum_{i=1}^m
 \big( \Omega(x_1-z_i,B_1|\tau) - \Omega(x_1-y_1,B_1|\tau) \big) \, c_i \notag \\
 &\quad + dx_1 \sum_{i=m+1}^n
 \big( \Omega(x_1-p_1,B_1|\tau) - \Omega(x_1-y_1,B_1|\tau) \big) \, c_i 
 \label{jncase1}
\end{align}
involving the generators $c_1,\cdots,c_n$ associated with the additional punctures $z_1,\cdots,z_n$
on both surfaces. If $x=x_1 \in \Sigma _1 \setminus \mD_1$ and $y=y_2 \in \Sigma _2 \setminus \mD_2$
are chosen to be on different surfaces, however, the generators $c_{m+1},\cdots,c_n$ associated
with the punctures $z_{m+1},\cdots,z_n \in \Sigma _2 \setminus \mD_2$ on a different surface than $x_1$
are absent,
\begin{align}
\cJ_n(z_1,\cdots,z_n;x_1, y_2 ) &\to dx_1 \sum_{i=1}^m
 \big( \Omega(x_1-z_i,B_1|\tau) - \Omega(x_1-p_1,B_1|\tau) \big) \, c_i 
 \label{jncase2}
\end{align}
\end{enumerate}

\newpage

\section{Conclusions and further directions}
\label{sec:conc}

In this work, we have presented an explicit construction of polylogarithms on compact Riemann surfaces of arbitrary genus. Generalizing the approach of Brown and Levin for the genus-one case, our construction relies on a flat connection whose path-ordered exponential plays the role of a generating series for higher-genus polylogarithms and manifests their homotopy-invariance. The flat connection takes values in the freely-generated Lie algebra introduced by Enriquez and Zerbini in \cite{Enriquez:2021} and is assembled from convolutions on higher-genus surfaces of Arakelov Green functions and their derivatives. Our construction furnishes the first explicit proposal for  a ``complete'' set of integration kernels beyond genus one:  the higher-genus polylogarithms constructed here are conjectured to close under taking primitives with respect to the points on the surface.

\sm

While our construction of higher-genus polylogarithms builds on the Brown-Levin connection at genus one, it also draws heavily on the structure of families of higher-genus modular tensors.  We illustrate the importance of these
tensorial building blocks and their properties in several examples of higher-genus polylogarithms, introduce a basis of polylogarithms which themselves  transform as modular tensors and provide non-trivial evidence for their closure under taking primitives. Moreover, upon separating degeneration of the Riemann surface, our flat connection reduces to flat connections on the degeneration components. We illustrate this result here for the case of genus two, thereby paving the way for systematic investigations into the rich network of relations among the higher-genus polylogarithms that result from different types of degenerations. 

\sm

It is expected that the higher-genus polylogarithms constructed in this work will find broad applications in theoretical physics, ranging from multi-loop amplitudes in string theory to Feynman integrals in quantum field theory and beyond. At the same time, our results offer a concrete forward leap towards a coherent theory of integration on arbitrary Riemann surfaces and should prove relevant to questions in number theory and algebraic geometry. Among
the myriad of mathematical and physical open problems raised by our construction,  the following questions readily qualify for tractable follow-up research:

\sm

\begin{itemize}
\itemsep=-0in
\item[(i)] proving the conjecture advanced here that higher-genus polylogarithms close under taking primitives,
based on the generalizations to all orders of the interchange lemma and Fay identities discussed in section \ref{sec:close} which are provided and proven in \cite{DHoker:2024ozn};
\item[(ii)] obtaining the separating and non-separating degenerations of the polylogarithms for arbitrary genera, exploiting the properties of the Arakelov Green function in \cite{DHoker:2017pvk, DHoker:2018mys};
\item[(iii)] determining the differential relations with respect to moduli variations  satisfied by higher-genus polylogarithms and their integration kernels using complex-structure deformation theory  \cite{DHoker:2014oxd, Basu:2018bde};

\item[(iv)] identifying generalizations of the (non-holomorphic) higher-genus modular graph tensors in \cite{Kawazumi:lecture, Kawazumi:seminar, DHoker:2020uid} that close under complex-structure variations and degenerations; 
\item[(v)] exploring the properties of the higher-genus associators proposed in section \ref{sec:assoc}, thereby generalizing the studies of elliptic associators introduced in \cite{Calaque:2009, Enriquez:2014, Hain:2013};
\item[(vi)] re-formulation of higher-genus string amplitudes in terms of the integration kernels and polylogarithms constructed in this work, a program that was foreshadowed long ago by the cohomological analysis of chiral blocks  for the case of genus two in \cite{DHoker:2007csw} and implemented more recently at genus one in \cite{Broedel:2014vla, Gerken:2018,Broedel:2017jdo}. 
\end{itemize}
We plan to report progress on some of these topics in future work.

\newpage

\appendix

\section{Definiton of the prime form}
\label{sec:pform}
\setcounter{equation}{0}

Given that the Arakelov Green function is constructed from the prime form $E(x,y)$
in (\ref{polylog.25}) and (\ref{polylog.23}), we shall briefly review the definition of the prime form in
this appendix. For this purpose, we introduce the Riemann $\tet$-function of rank $h$,
\beq
\vartheta[\kappa](\zeta | \Omega) = \! \sum_{ n \in \mathbb Z^h } \!  
e^{ \pi i  (n{+}\kappa' )^t \Omega (n{+} \kappa')  + 2\pi i (n{+} \kappa' )^t (\zeta {+} \kappa'') } 
\label{appth.01}
\eeq
where $\zeta \in \mathbb C^h$ and where the characteristics $\kappa = [\kappa', \kappa'']$ 
comprises two $h$-component vectors $\kappa', \kappa'' \in \mathbb C^h$. Following the
main text, we will henceforth suppress the dependence on the period matrix~$\Omega$.

\sm

The prime form at arbitrary genus $h$
is built from a specialization of $\kappa$ to odd half-characteristics
or spin structures $\nu = [\nu', \nu'']$ with entries $\in \{ 0,\frac{1}{2} \}$ 
such that $4 \nu' \cdot \nu''$ is odd~\cite{Fay:1973}:
\beq
E(x,y   ) = \frac{ \vartheta[\nu]\big( \int^x_y \bom  \big)}{h_\nu(x ) h_\nu(y ) } 
\label{appth.02}
\eeq
By virtue of the holomorphic $(\frac{1}{2},0)$-forms $h_\nu(x ) $ in the denominator subject to
\beq
h_\nu(x)^2 = \sum_{I=1}^h \bom_I(x) \frac{\partial}{\partial \zeta_I} \vartheta[\nu](\zeta) \, \Big|_{\zeta = 0}
\eeq
the definition (\ref{appth.02}) of the prime form is independent on the choice of
the odd spin structure~$\nu$. Since the $\tet$-functions entering the prime form 
are odd under the flip $\zeta \rightarrow - \zeta$ in (\ref{appth.01}), the prime form is antisymmetric
$E( y, x   ) =  - E(x,y )$ under exchange of $x,y$ and in particular exhibits
the short-distance behaviour
$ E(x,y )= (x-y) + {\cal O}\big(  (x-y)^3\big)$ in local complex coordinates.

\newpage

\section{Separating degeneration of $\p_x \cG^I(x,y)$}
\label{sec:sepapp}
\setcounter{equation}{0}

In this appendix, we derive the separating degeneration of the functions $\p_x \cG^I(x,y)$ for genus two with $I=1,2$. These derivations are essential components of Lemma \ref{6.thm3}. Actually, it will be convenient to derive first the asymptotic behavior in the separating degeneration of the modular tensor with a lowered index, defined by $\p_x \cG_I(x,y) = Y_{IJ} \, \p_x \cG^J(x,y)$, and then reconvert the result to $\p_x\cG^J(x,y)$.

\subsection{Degeneration of $\p_{x_1} \cG_1(x_1,y_1)$ and $\p_{x_2} \cG_2(x_2,y_2)$}

Following the decomposition of (\ref{6.decomp}), we evaluate the contributions $\cG^{(1)}_1(x,y)$ and $\cG^{(2)}_2(x,y)$ by substituting the degeneration limits of the various factors in its integrand,
\bea
\cG_1^{(1)}(x_1,y_1) & = & \int  _{\Sigma_1 \setminus \mD_1}  \!\!\!\!\!  d^2 z_1 \, 
\Big [ g(x_1-z_1|\tau) - \thalf g(x_1-p_1|\tau) - \thalf g(z_1-p_1|\tau) - \thalf \ln |\hat v | \Big ]  
\no \\ && \hskip 0.7in \times
 \Big [ \p_{z_1}  g(z_1-y_1|\tau) - \thalf \p_{z_1} g(z_1-p_1|\tau) \Big ] 
\eea
The integral may be extended to all of $\Sigma_1$ where it  is absolutely convergent and differs from the original integral by a contribution that is proportional to the coordinate volume, which is of order $\cO(v)$ in view of (\ref{6.dist}),  and will be neglected. As an integral over $\Sigma_1$, the contributions of the $z_1$-independent terms inside the first bracket vanish in view of,
\bea
 \int  _{\Sigma_1} d^2 z_1 \, \p_{z_1}  \Big( g(z_1-y_1|\tau) \Big)^n  =0 \hskip 1in n \geq 0
 \eea
Evaluating the remaining integrals using the concatenated Green functions $g_n$ of (\ref{2.concat}), and taking the derivative in $x_1$, gives the following limit,
\bea
\p_{x_1} \cG_1^{(1)} (x_1, y_1) \to 
(\Im \tau ) \Big ( \p_{x_1}^2 g_2(x_1-y_1|\tau) - \thalf \p_{x_1}^2 g_2 (x_1-p_1|\tau) \Big )
 \label{eqapp.03}
\eea
The overall factor of $\Im \tau$ arises from the integrand in the definition of $g_2$ in (\ref{2.concat}).  
 
\sm
 
We proceed analogously for $ \p_{x_1} \cG_I^{(2)}(x_1,y_1)$, 
\bea
\cG_1^{(2)}(x_1,y_1) & = & { 1 \over 4} \overline{\left ( { v \over 2 \pi i \hat \om_2(p_2)} \right )}
\int  _{\Sigma_2 \setminus \mD_2}  \!\!\!\!\! d^2 z_2 \, \Big [g(x_1-p_1|\tau) + g(z_2-p_2|\sigma) + \ln |\hat v| \Big ] 
\no \\ && \hskip 1.5in
\times  \overline{ \p_{z_2} \p_{p_2} \ln \tet_1(z_2-p_2|\sigma)}\, 
\p_{z_2}  g(z_2-p_2|\sigma) 
\eea
On the face of it, this contribution is automatically suppressed by a factor of $v$. However, the integral does not extend to a convergent integral over $\Sigma _1$, so part or all of the suppression might in principle be cancelled. To proceed, we extract the leading singularity near the puncture in terms of a contour integral over $\p \mD_1$ which may be evaluated exactly in the limit where the size of the disc $\mD_1$ is small. First, we show that the combination, 
\bea
\cI= \int  _{\Sigma_2 \setminus \mD_2}  \!\!\!\!\! d^2 z_2 \, 
 \overline{ \p_{z_2} \p_{p_2} \ln \tet_1(z_2-p_2|\sigma)}\, 
\p_{z_2}  g(z_2-p_2|\sigma) 
\eea
remains bounded as $v\to 0$ so that its contribution to $\cG_1^{(2)}(x_1,y_1)$ vanishes in this limit. To do so we write it as follows,
\bea
\cI & = & - \int  _{\Sigma_2 \setminus \mD_2}  \!\!\!\!\! d^2 z_2 \, 
 \overline{ \p_{p_2} \ln \tet_1(z_2-p_2|\sigma)}\, 
\p_{\bar z_2} \p_{z_2}  g(z_2-p_2|\sigma) 
\no \\ &&
- { i \over 2} \int  _{\Sigma_2 \setminus \mD_2}  d \Big [ dz_2 \, \p_{z_2} g(z_2-p_2|\sigma)
 \overline{ \p_{p_2} \ln \tet_1(z_2-p_2|\sigma)} \Big ]
 \label{eqapp.06}
\eea
Using $\p_{\bar z_2} \p_{z_2}  g(z_2-p_2|\sigma) = - \pi \delta(z_2,p_2) + \frac{ \pi}{\Im \sigma}$, the fact that the support of the $\delta$-functions is outside the domain of integration, and that the integral involving the constant term is convergent on $\Sigma_1$ and integrates to zero there, we see that the first line on the right vanishes. The remainder may be written as follows, 
\bea
\cI =  - { i \over 2} \oint  _{\p \mD_2}  dz_2 \, \big | \p_{z_2} g(z_2-p_2|\sigma) \big |^2 
 \label{eqapp.07}
\eea
In passing from the second line of (\ref{eqapp.06}) to (\ref{eqapp.07}) we have omitted a contribution proportional to $2\pi i  \Im(z_2-p_2) / \Im \sigma$, as the integral involving this term is of order $\cO(v)$.
The overall minus sign results from the fact that the orientation of the integrations over  the boundary of $\Sigma_2 \setminus \mD_2$ is opposite to the ones over the boundary of $\mD_2$.  The angular integration of the pole term vanishes, and the remaining contribution is bounded as $v\to 0$.  We conclude that $\cG^{(2)}_1(x_1,y_1) \to 0$ 
so that we obtain the following limit for $\p_{x_1} \cG_1(x_1, y_1)$ from (\ref{eqapp.03}) and, by swapping the role of the two surfaces,  
\bea
\label{A.limi1}
\p_{x_1} \cG_1(x_1, y_1) & \to &
(\Im \tau ) \Big ( \p_{x_1}^2 g_2(x_1-y_1|\tau) - \thalf \p_{x_1}^2 g_2 (x_1-p_1|\tau) \Big )
\no \\
\p_{x_2} \cG_2 (x_2, y_2) & \to &
(\Im \sigma ) \Big ( \p_{x_2}^2 g_2(x_2-y_2|\sigma) - \thalf \p_{x_2}^2 g_2 (x_2-p_2|\sigma) \Big )
\eea
where we have neglected all contributions that vanish as $v \to 0$.

\subsection{Degeneration of $\p_{x_1} \cG_2(x_1,y_1)$ and $\p_{x_2} \cG_1(x_2,y_2)$}

Following the corresponding analysis for these two cases, we find the limits,
\bea
\label{A.limits2}
\p_{x_1} \cG_2(x_1,y_1) & \to & 0
\no \\
\p_{x_2} \cG_1(x_2,y_2) & \to & 0
\eea

\subsection{Degeneration of $\p_{x_1} \cG_1(x_1,y_2)$ and $\p_{x_2} \cG_2(x_2,y_1)$}
\label{sec:appa.3}

Using once more the decomposition of (\ref{6.decomp}), we have,
\bea
\cG_1^{(1)}(x_1,y_2) & = & \half \int  _{\Sigma_1 \setminus \mD_1}  \!\!\!\!\! d^2 z_1 \, 
\Big [ g(x_1-z_1|\tau) - \thalf g(x_1-p_1|\tau) - \thalf g(z_1-p_1|\tau) - \thalf \ln |\hat v | \Big ]  
\no \\ && \hskip 0.7in \times
 \p_{z_1}  g(z_1-p_1|\tau)
\eea
The integrals may be continued to absolutely convergent integrals on $\Sigma_1$. The $z_1$-independent terms inside the  bracket  integrate to zero, as does the term $g(z_1-p_1|\tau)$ so that,
\bea
\! \! \! \cG_1^{(1)}(x_1,y_2) & = & \half \int_{\Sigma_1}   \! d^2 z_1 \, 
g(x_1-z_1|\tau) \,  \p_{z_1}  g(z_1-p_1|\tau) 
\no \\ & = & \half (\Im \tau) \p_{x_1} g_2(x_1-p_1|\tau)
\label{nzeropart}
\eea
The remaining contribution is given by,
\bea
\cG_1^{(2)}(x_1,y_2) & = & { 1 \over 2} \overline{\left ( { v \over 2 \pi i \hat \om_2(p_2)} \right )} 
\int  _{\Sigma_2 \setminus \mD_2}  \!\!\!\!\! d^2 z_2 \, 
\Big [ g(x_1-p_1|\tau) + g(z_2-p_2|\sigma) + \ln |\hat v| \Big ] 
\no \\ && \hskip 0.3in
\times  \overline{ \p_{z_2} \p_{p_2} \ln \tet_1(z_2-p_2|\sigma)}\, 
\Big [ \p_{z_2}  g(z_2-y_2|\sigma) - \thalf \p_{z_2} g(z_2-p_2|\sigma) \Big ]
\eea
The contribution of the second term in the bracket on the second line may be evaluated by the method used for $\cG^{(2)}_1(x_1,y_1)$ and vanishes as $v \to 0$. In the remaining expression the $z_2$-independent terms in the first bracket integrate to a finite quantity as $v \to 0$, leaving contributions proportional to,
\bea
\cG_1^{(2)}(x_1,y_2) & \sim & \bar v
\int  _{\Sigma_2 \setminus \mD_2}  \!\!\!\!\! d^2 z_2 \, 
\overline{ \p_{z_2} \p_{p_2} \ln \tet_1(z_2-p_2|\sigma)}\, 
\p_{z_2}  g(z_2-y_2|\sigma) g(z_2-p_2|\sigma) 
\eea
Isolating the leading singularity in the form of a contour integral,
\bea
\cG_1^{(2)}(x_1,y_2) & \sim &
- \bar v
\int  _{\Sigma_2 \setminus \mD_2}  \!\!\!\!\! d^2 z_2 \, 
\overline{ \p_{p_2} \ln \tet_1(z_2-p_2|\sigma)}\, 
\p_{\bar z_2} \Big [ \p_{z_2}  g(z_2-y_2|\sigma) g(z_2-p_2|\sigma) \Big ]
\no \\ &&
+ \frac{ i}{ 2} \bar v
\oint  _{\p \mD_2} \,  dz_2 \,  \p_{z_2}  g(z_2-y_2|\sigma) g(z_2-p_2|\sigma) 
\overline{ \p_{p_2} \ln \tet_1(z_2-p_2|\sigma)}
\eea
The angular integration of the second term cancels the pole and gives a finite result which vanishes as $v \to 0$. Applying $\p_{\bar z_2}$ to the first factor in the bracket of the first line gives a $\delta(z_2,y_2)$ and a constant term, both of which give finite contributions to the integral leading to vanishing contributions to $\cG_1^{(2)}(x_1,y_2) $ as $v \to 0$. The remaining integral is then, 
\bea
\cG_1^{(2)}(x_1,y_2) & \sim &
- \bar v  \int  _{\Sigma_2 \setminus \mD_2}  \!\!\!\!\! d^2 z_2 \, 
\overline{ \p_{p_2} \ln \tet_1(z_2-p_2|\sigma)}\, 
 \p_{z_2}  g(z_2-y_2|\sigma) \p_{\bar z_2}  g(z_2-p_2|\sigma) 
\eea
The angular integration again kills the pole and we are left with a finite integral with a vanishing contribution as $v \to 0$. Hence, by the vanishing asymptotics of $\cG_1^{(2)}(x_1,y_2) $ and the non-zero contribution (\ref{nzeropart}) from $\cG_1^{(1)}(x_1,y_2) $, we find the following asymptotics for $\p_{x_1} \cG_1(x_1,y_2)$ and, by swapping the two surfaces, we obtain,
\bea
\label{A.limits3}
\p_{x_1} \cG_1(x_1,y_2) & \to & \thalf (\Im \tau) \, \p_{x_1}^2 g_2(x_1-p_1|\tau)
\no \\
\p_{x_2} \cG_2(x_2,y_1) & \to & \thalf (\Im \sigma) \, \p_{x_2}^2 g_2(x_2-p_2|\sigma)
\eea

\subsection{Degeneration of $\p_x \cG^I(x,y)$}

Assembling the results obtained in (\ref{A.limi1}), (\ref{A.limits2}), and (\ref{A.limits3}) for $\p_x \cG_J(x,y)$, we obtain those of $\cG^I(x,y)$ by raising the index $J$ with the help of $Y^{IJ}$. The off-diagonal elements of $Y^{IJ}$ are proportional to $v$ and do not contribute in the limit $v \to 0$. The role of the diagonal elements of $Y^{IJ}$ is to cancel the prefactor $\Im \tau$ in (\ref{A.limi1}) and $\Im \sigma$ in (\ref{A.limits3}). The final result is the following table of limits,
\bea
\label{A.limit}
\p_{x_1} \cG^1(x_1, y_1) & \to &
\p_{x_1}^2 g_2(x_1-y_1|\tau) - \thalf \p_{x_1}^2 g_2 (x_1-p_1|\tau) 
\no \\
\p_{x_1} \cG^1(x_1,y_2) & \to & \thalf  \p_{x_1}^2 g_2(x_1-p_1|\tau)
\no \\
\p_{x_1} \cG^2(x_1,y_1) & \to & 0
\no \\
\no \\
\p_{x_2} \cG^2 (x_2, y_2) & \to &
\p_{x_2}^2 g_2(x_2-y_2|\sigma) - \thalf \p_{x_2}^2 g_2 (x_2-p_2|\sigma) 
\no \\
\p_{x_2} \cG^2(x_2,y_1) & \to & \thalf  \p_{x_2}^2 g_2(x_2-p_2|\sigma)
\no \\
\p_{x_2} \cG^1(x_2,y_2) & \to & 0
\eea
Together with $\p_{x}^2 g_2(x{-}p|\tau) = - f^{(2)}(x{-}p|\tau)$ as in (\ref{2.fg}),
this completes the proof of Lemma~\ref{6.thm3} for the case $s=1$.



\newpage


\begin{thebibliography}{99}
\itemsep=-0.02in

\bibitem{Bourjaily:2022bwx}
J.~L.~Bourjaily, \textit{et al.}
``Functions Beyond Multiple Polylogarithms for Precision Collider Physics,''
[arXiv:2203.07088 [hep-ph]].

\bibitem{Abreu:2022mfk}
S.~Abreu, R.~Britto and C.~Duhr,
``The SAGEX review on scattering amplitudes Chapter 3: Mathematical structures in Feynman integrals,''
J. Phys. A \textbf{55} (2022) no.44, 443004
[arXiv:2203.13014].

\bibitem{Blumlein:2022qci}
J.~Bl\"umlein and C.~Schneider,
``The SAGEX review on scattering amplitudes Chapter 4: Multi-loop Feynman integrals,''
J. Phys. A \textbf{55} (2022) no.44, 443005
[arXiv:2203.13015].

\bibitem{Weinzierl:2022}
S.~Weinzierl, ``Feynman Integrals,
a Comprehensive Treatment for Students and Researchers,'' 
Springer Cham (2022), ISBN
978-3-030-99557-7.

\bibitem{Berkovits:2022ivl}
N.~Berkovits, E.~D'Hoker, M.~B.~Green, H.~Johansson and O.~Schlotterer,
``Snowmass White Paper: String Perturbation Theory,''
[arXiv:2203.09099].

\bibitem{Dorigoni:2022iem}
D.~Dorigoni, M.~B.~Green and C.~Wen,
``The SAGEX review on scattering amplitudes Chapter 10: Selected topics on modular covariance of type IIB string amplitudes and their~~supersymmetric Yang\textendash{}Mills duals,''
J. Phys. A \textbf{55} (2022) no.44, 443011
[arXiv:2203.13021].

\bibitem{DHoker:2022dxx}
E.~D'Hoker and J.~Kaidi,
``Lectures on modular forms and strings,''
[arXiv:2208.07242].


\bibitem{Mafra:2022wml}
C.~R.~Mafra and O.~Schlotterer,
``Tree-level amplitudes from the pure spinor superstring,''
Phys. Rept. \textbf{1020} (2023), 1-162
[arXiv:2210.14241].

\bibitem{DK}
E.~D'Hoker and J.~Kaidi, ``Modular forms and string theory,'' Cambridge University Press (2024), ISBN 978-1-009-45753-8.


\bibitem{Goncharov:1995}
A.~B.~Goncharov,
``Geometry of Configurations, Polylogarithms, and Motivic Cohomology,''
Advances in Mathematics
{\bf 114} (1995), 197-318.

\bibitem{Goncharov:1998kja}
A.~B.~Goncharov,
``Multiple polylogarithms, cyclotomy and modular complexes,''
Math. Res. Lett. \textbf{5} (1998), 497-516
[arXiv:1105.2076 [math.AG]].

\bibitem{Goncharov:2001iea}
A.~B.~Goncharov,
``Multiple polylogarithms and mixed Tate motives,''
[arXiv:math/0103059 [math.AG]].

\bibitem{Remiddi:1999ew}
E.~Remiddi and J.~A.~M.~Vermaseren,
``Harmonic polylogarithms,''
Int. J. Mod. Phys. A \textbf{15} (2000), 725-754
[arXiv:hep-ph/9905237].

\bibitem{Vollinga:2004sn}
J.~Vollinga and S.~Weinzierl,
``Numerical evaluation of multiple polylogarithms,''
Comput. Phys. Commun. \textbf{167} (2005), 177
[arXiv:hep-ph/0410259].

\bibitem{Goncharov:2010jf}
A.~B.~Goncharov, M.~Spradlin, C.~Vergu and A.~Volovich,
``Classical Polylogarithms for Amplitudes and Wilson Loops,''
Phys. Rev. Lett. \textbf{105} (2010), 151605
[arXiv:1006.5703].

\bibitem{Duhr:2012fh}
C.~Duhr,
``Hopf algebras, coproducts and symbols: an application to Higgs boson amplitudes,''
JHEP \textbf{08} (2012), 043
[arXiv:1203.0454].

\bibitem{Broedel:2013tta}
J.~Broedel, O.~Schlotterer and S.~Stieberger,
``Polylogarithms, Multiple Zeta Values and Superstring Amplitudes,''
Fortsch. Phys. \textbf{61} (2013), 812-870
[arXiv:1304.7267].

\bibitem{Schlotterer:2018zce}
O.~Schlotterer and O.~Schnetz,
``Closed strings as single-valued open strings: A genus-zero derivation,''
J. Phys. A \textbf{52} (2019) no.4, 045401
[arXiv:1808.00713].

\bibitem{Vanhove:2018elu}
P.~Vanhove and F.~Zerbini,
``Single-valued hyperlogarithms, correlation functions and closed string amplitudes,''
Adv.\ Theor.\ Math.\ Phys.\ {\bf 26} (2022) no.2, 455-530  [arXiv:1812.03018].


\bibitem{Beilinson:1994}
A.~Beilinson and A.~Levin, 
``The Elliptic Polylogarithm," 
in Proc.\ of Symp.\ in Pure Math.\ {\bf 55} (1994), 
123-190.

\bibitem{Levin:2007}
A.~Levin and G.~Racinet, ``Towards multiple elliptic polylogarithms," 
[arXiv:math/0703237 [math.NT]].

\bibitem{BrownLevin}
F.~Brown and A.~Levin, ``Multiple Elliptic Polylogarithms,'' [arXiv:1110.6917 [math.NT]].


\bibitem{Bloch:2013tra}
S.~Bloch and P.~Vanhove,
``The elliptic dilogarithm for the sunset graph,''
J. Number Theor. \textbf{148} (2015), 328-364
[arXiv:1309.5865].

\bibitem{Adams:2017ejb}
L.~Adams and S.~Weinzierl,
``Feynman integrals and iterated integrals of modular forms,''
Commun. Num. Theor. Phys. \textbf{12} (2018), 193-251
[arXiv:1704.08895].

\bibitem{Ablinger:2017bjx}
J.~Ablinger, J.~Bl\"umlein, A.~De Freitas, M.~van Hoeij, E.~Imamoglu, C.~G.~Raab, C.~S.~Radu and C.~Schneider,
``Iterated Elliptic and Hypergeometric Integrals for Feynman Diagrams,''
J. Math. Phys. \textbf{59} (2018) no.6, 062305
[arXiv:1706.01299].

\bibitem{Broedel:2017kkb}
J.~Broedel, C.~Duhr, F.~Dulat and L.~Tancredi,
``Elliptic polylogarithms and iterated integrals on elliptic curves. Part I: general formalism,''
JHEP \textbf{05} (2018), 093
[arXiv:1712.07089].


\bibitem{Broedel:2014vla}
J.~Broedel, C.~R.~Mafra, N.~Matthes and O.~Schlotterer,
``Elliptic multiple zeta values and one-loop superstring amplitudes,''
JHEP \textbf{07} (2015), 112
[arXiv:1412.5535].


\bibitem{DHoker:2015wxz}
E.~D'Hoker, M.~B.~Green, \"O.~G\"urdogan and P.~Vanhove,
``Modular Graph Functions,''
Commun. Num. Theor. Phys. \textbf{11} (2017), 165-218
[arXiv:1512.06779].

\bibitem{Dolan:2007eh}
L.~Dolan and P.~Goddard,
``Current Algebra on the Torus,''
Commun. Math. Phys. \textbf{285} (2009), 219-264
[arXiv:0710.3743].

\bibitem{Tsuchiya:2017joo}
A.~G.~Tsuchiya,
``On new theta identities of fermion correlation functions on genus g Riemann surfaces,''
[arXiv:1710.00206].

\bibitem{Gerken:2018}
J.~E. Gerken, A.~Kleinschmidt and O.~Schlotterer,
``Heterotic-string amplitudes at one loop: modular graph forms and relations to open strings,''
JHEP {\bf 01} (2019), 052
[arXiv:1811.02548].




\bibitem{Green:1987mn}
M.~B.~Green, J.~H.~Schwarz and E.~Witten,
``Superstring Theory Vol.\ 2: Loop Amplitudes, Anomalies and Phenomenology,''
Cambridge University Press (1988),\\
ISBN 978-0-521-35753-1.

\bibitem{DHoker:1988pdl}
E.~D'Hoker and D.~H.~Phong,
``The Geometry of String Perturbation Theory,''\\
Rev. Mod. Phys. \textbf{60} (1988), 917.

\bibitem{Pol}
J.~Polchinski, ``String Theory Vol II,'' Cambridge University Press (1997), \\ ISBN 978-0-521-63304-8.

\bibitem{Witten:2012bh}
E.~Witten,
``Superstring Perturbation Theory Revisited,''
[arXiv:1209.5461].

\bibitem{Huang:2013kh}
R.~Huang and Y.~Zhang,
``On Genera of Curves from High-loop Generalized Unitarity Cuts,''
JHEP \textbf{04} (2013), 080
[arXiv:1302.1023 [hep-ph]].

\bibitem{Georgoudis:2015hca}
A.~Georgoudis and Y.~Zhang,
``Two-loop Integral Reduction from Elliptic and Hyperelliptic Curves,''
JHEP \textbf{12} (2015), 086
[arXiv:1507.06310].

\bibitem{Doran:2023yzu}
C.~F.~Doran, A.~Harder, E.~Pichon-Pharabod and P.~Vanhove,
``Motivic geometry of two-loop Feynman integrals,''
Quart. J. Math. Oxford Ser. \textbf{75} (2024) no.3, 901-967
[arXiv:2302.14840 [math.AG]].

\bibitem{Brown:2010bw}
F.~Brown and O.~Schnetz,
``A K3 in $\phi^4$,''
Duke Math. J. \textbf{161} (2012) no.10, 1817-1862
[arXiv:1006.4064 [math.AG]].

\bibitem{Bourjaily:2018yfy}
J.~L.~Bourjaily, A.~J.~McLeod, M.~von Hippel and M.~Wilhelm,
``Bounded Collection of Feynman Integral Calabi-Yau Geometries,''
Phys. Rev. Lett. \textbf{122} (2019) no.3, 031601
[arXiv:1810.07689].

\bibitem{Duhr:2022pch}
C.~Duhr, A.~Klemm, F.~Loebbert, C.~Nega and F.~Porkert,
``Yangian-Invariant Fishnet Integrals in Two Dimensions as Volumes of Calabi-Yau Varieties,''
Phys. Rev. Lett. \textbf{130} (2023) no.4, 4
[arXiv:2209.05291].

\bibitem{Pogel:2022vat}
S.~P\"ogel, X.~Wang and S.~Weinzierl,
``Bananas of equal mass: any loop, any order in the dimensional regularisation parameter,''
JHEP \textbf{04} (2023), 117
[arXiv:2212.08908].

\bibitem{Duhr:2022dxb}
C.~Duhr, A.~Klemm, C.~Nega and L.~Tancredi,
``The ice cone family and iterated integrals for Calabi-Yau varieties,''
JHEP \textbf{02} (2023), 228
[arXiv:2212.09550].

\bibitem{Enriquez:2011}
B.~Enriquez, 
``Flat connections on configuration spaces and braid groups of surfaces,"
Advances in Mathematics {\bf 252} (2014), 204--226
[arXiv:1112.0864 [math.GT]].

\bibitem{Enriquez:2021}
B.~Enriquez and F.~Zerbini,
``Construction of Maurer-Cartan elements over configuration spaces of curves,''
[arXiv:2110.09341 [math.AG]].

\bibitem{Enriquez:2022}
B.~Enriquez and F.~Zerbini,
``Analogues of hyperlogarithm functions on affine complex curves,''
[arXiv:2212.03119 [math.AG]].


\bibitem{Bernard:1988}
D.~Bernard,
``On the Wess-Zumino-Witten models on Riemann surfaces,"
Nucl. Phys. B {\bf 309} (1988), 145-174.



\bibitem{Faltings}
G.~Faltings, ``Calculus on Arithmetic Surfaces,'' Ann. Math. {\bf 119} (1984) 387.

\bibitem{Alvarez-Gaume:1986nqf}
L.~Alvarez-Gaume, G.~W.~Moore, P.~C.~Nelson, C.~Vafa and J.~b.~Bost,
``Bosonization in Arbitrary Genus,''
Phys. Lett. B \textbf{178}, 41-47 (1986).

\bibitem{Fay:1973}
J.~D.~Fay, ``Theta Functions on Riemann Surfaces,'' Lecture Notes in Math. {\bf 352} (1973).

\bibitem{DHoker:2017pvk}
E.~D'Hoker, M.~B.~Green and B.~Pioline,
``Higher genus modular graph functions, string invariants, and their exact asymptotics,''
Commun. Math. Phys. \textbf{366} (2019) 927-979
[arXiv:1712.06135].

\bibitem{DHoker:2013fcx}
E.~D'Hoker and M.~B.~Green,
``Zhang-Kawazumi Invariants and Superstring Amplitudes,''
J. Number Theor. \textbf{144} (2014), 111-150
[arXiv:1308.4597].

\bibitem{DHoker:2014oxd}
E.~D'Hoker, M.~B.~Green, B.~Pioline and R.~Russo,
``Matching the $D^{6}R^{4}$ interaction at two-loops,''
JHEP \textbf{01} (2015), 031
[arXiv:1405.6226].

\bibitem{DHoker:2018mys}
E.~D'Hoker, M.~B.~Green and B.~Pioline,
``Asymptotics of the $D^8 R^4$ genus-two string invariant,''
Commun. Num. Theor. Phys. \textbf{13} (2019), 351-462
[arXiv:1806.02691].

\bibitem{DHoker:2020tcq}
E.~D'Hoker, C.~R.~Mafra, B.~Pioline and O.~Schlotterer,
``Two-loop superstring five-point amplitudes. Part II. Low energy expansion and S-duality,''
JHEP \textbf{02} (2021), 139
[arXiv:2008.08687].

\bibitem{Kawazumi:lecture}
N.~Kawazumi, 
``Some tensor field on the Teichm\"uller space,''
\href{http://www.ms.u-tokyo.ac.jp/~kawazumi/OIST1610_v1.pdf}{Lecture at MCM2016}, OIST (2016).

\bibitem{Kawazumi:seminar}
N.~Kawazumi, 
``Differential forms and functions on the moduli space of Riemann surfaces," in S\'eminaire Alg\`ebre et topologie, Universit\'e de Strasbourg. 
\href{http://www.ms.u-tokyo.ac.jp/~kawazumi/1701Strasbourg_v1.pdf}{S\'eminaire Alg\`ebre et topologie, Universit\'e de Strasbourg} (2017).

\bibitem{DHoker:2020uid}
E.~D'Hoker and O.~Schlotterer,
``Identities among higher genus modular graph tensors,''
Commun. Num. Theor. Phys. \textbf{16} (2022) no.1, 35-74
[arXiv:2010.00924].

\bibitem{Kawazumi:paper}
N.~Kawazumi, ``A twisted invariant of a compact Riemann surface,'' [arXiv:2210.00532 [math.GT]].

  \bibitem{vdG2}
F.~Cl\'ery, G.~van der Geer, S.~Grushevsky,
 ``Siegel modular forms of genus 2 and level~2,"
[arXiv:1306.6018 [math.AG]].
 
   \bibitem{vdG3}
F.~Cl\'ery, G.~van der Geer,
 ``Constructing vector-valued Siegel modular forms from scalar-valued Siegel modular forms,"
[arXiv:1409.7176 [math.AG]].
 
  \bibitem{vdG4}
 G.~van der Geer,
 ``Siegel modular forms of degree two and three and invariant theory,"
[arXiv:2102.02245 [math.AG]].
 
\bibitem{Panzer:2015ida}
E.~Panzer, ``Feynman integrals and hyperlogarithms,''
[arXiv:1506.07243 [math-ph]].


\bibitem{Broedel:2018iwv}
J.~Broedel, C.~Duhr, F.~Dulat, B.~Penante and L.~Tancredi,
``Elliptic symbol calculus: from elliptic polylogarithms to iterated integrals of Eisenstein series,''
JHEP \textbf{08} (2018), 014
[arXiv:1803.10256].

\bibitem{Mumford:1983}
D.~Mumford, ``Tata Lectures on Theta I,'' Progress in Mathematics {\bf 28} (1983), Birkh\"auser Boston Inc., MA.
 
 \bibitem{Reut}
C.~Reutenauer, ``Free Lie Algebras", Ocford University Press (1993).

\bibitem{Deligne:1989}
P.~Deligne,
``Le Groupe Fondamental de la Droite Projective Moins Trois Points,''
in Galois Groups Over, vol. 16, ed. by Y.~Ihara, K.~Ribet, J.P.~Serre.
Mathematical Sciences Research Institute Publications (Springer, New York, 1989), 79-297.

\bibitem{Brown:2014}
F.~Brown,
``Motivic periods and the projective line minus three points,''
in Proceedings of the ICM 2014, vol. 2, ed. by S.Y.~Jang Y.R.~Kim, D.-W.~Lee, I.~Yie (2014), 295-318.

\bibitem{DHoker:2024ozn}
E.~D'Hoker and O.~Schlotterer,
``Fay identities for polylogarithms on higher-genus Riemann surfaces,''
[arXiv:2407.11476].

\bibitem{Drinfeld:1989}
V.~Drinfeld, ``Quasi Hopf algebras,'' Leningrad Math. {\bf J. 1} (1989), 1419.

\bibitem{Drinfeld:1991}
V.~Drinfeld, ``On quasitriangular quasi-Hopf algebras and on a group that is closely
connected with Gal($\bar{\mathbb Q}/\mathbb Q$),'' Leningrad Math. {\bf J. 2} (1991), 829.

\bibitem{Calaque:2009}
D.~Calaque, B.~Enriquez and P.~Etingof, 
``Universal KZB equations: the elliptic case,''
Progr. Math. {\bf 269} (2009), 165-266. Birkh\"auser Boston Inc., MA.

\bibitem{Enriquez:2014}
B.~Enriquez, ``Elliptic associators,'' 
Selecta Math. (N.S.) {\bf 20} (2014), 491-584.

\bibitem{Hain:2013}
R.~Hain,
``Notes on the universal elliptic KZB equation,''
 [arXiv:1309.0580 [math.AG]]. 
 
 \bibitem{Gonzalez:2020}
M.~Gonzalez, ``Surface Drinfeld Torsors I: Higher Genus Associators,''
[arXiv:2004.07303 [math.QA]].
 
\bibitem{Basu:2018bde}
A.~Basu,
``Eigenvalue equation for genus two modular graphs,''
JHEP \textbf{02} (2019), 046
[arXiv:1812.00389].

\bibitem{DHoker:2007csw}
E.~D'Hoker and D.~H.~Phong,
``Two-Loop Superstrings. VII. Cohomology of Chiral Amplitudes,''
Nucl. Phys. B \textbf{804} (2008), 421-506
[arXiv:0711.4314].

\bibitem{Broedel:2017jdo}
J.~Broedel, N.~Matthes, G.~Richter and O.~Schlotterer,
``Twisted elliptic multiple zeta values and non-planar one-loop open-string amplitudes,''
J. Phys. A \textbf{51} (2018) no.28, 285401
[arXiv:1704.03449].


\end{thebibliography}
\end{document}